\documentclass[11pt,showpacs,preprintnumbers,superscriptaddress,amsmath,amssymb,nofootinbib]{revtex4}
\usepackage{graphicx}
\usepackage{dcolumn}
\usepackage{bm}
\usepackage{amssymb}
\usepackage{amsmath}
\usepackage{epsfig}    
\usepackage{color}
\usepackage{slashed}
\usepackage{hhline}
\usepackage{ulem}

\def\be{\begin{equation}}
\def\ee{\end{equation}}
\newcommand{\bea}{\begin{eqnarray}}
\newcommand{\eea}{\end{eqnarray}}
\newcommand{\nn}{\nonumber}

\begin{document}


\title{Radiative lepton seesaw model in a non-invertible fusion rule \\ and\\ gauged $B-L$ symmetry}


\author{Takaaki Nomura}
\email{nomura@scu.edu.cn}
\affiliation{College of Physics, Sichuan University, Chengdu 610065, China}

\author{Hiroshi Okada}
\email{hiroshi3okada@htu.edu.cn}
\affiliation{Department of Physics, Henan Normal University, Xinxiang 453007, China}

\date{\today}

\begin{abstract}
{
We propose a radiative lepton seesaw model with the ${\mathbb Z}_3$ Tambara-Yamagami fusion rule and gauged $B-L$ symmetry.
The heavier right-handed neutral fermions ($N_R$) simultaneously contribute to the masses of charged-leptons and neutrinos, and three families of $N_R$ are necessary in realizing realistic mass matrices.
On the other hand, we introduce a gauged $B-L$ symmetry to preserve the lepton seesaw mechanism. 
As a result, this gauged symmetry not only assures the three families of $N_R$ from view point of chiral gauge anomaly cancellation, but also provides a dominant annihilation cross section via Higgs portal in order to explain the correct relic density of dark matter. 
Finally, we show several numerical results satisfying the neutrino oscillation data, lepton flavor violations, muon $g-2$, and relic density of dark matter.
}
 %
 \end{abstract}
\maketitle
\newpage

\section{Introduction}

A symmetry is a key ingredient in describing high energy physics.
Recently, intriguing symmetries have been applied to particle physics in refs~\cite{Choi:2022jqy,Cordova:2022fhg,Cordova:2022ieu,Cordova:2024ypu,Kobayashi:2024cvp,Kobayashi:2024yqq,Kobayashi:2025znw,Suzuki:2025oov,Liang:2025dkm,Kobayashi:2025ldi, Kobayashi:2025lar}. 
For example, zero textures in the Yukawa matrix have been discussed in refs.~\cite{Branco:1988iq, Kobayashi:2024cvp,Kobayashi:2025znw}.
The masses and mixings in addition to CP phases have been studied by \cite{Kobayashi:2024cvp,Kobayashi:2025znw} in the quark sector and \cite{Kobayashi:2025ldi} for the lepton sector. 
%
%
{However, these models would be replaced by some group symmetries~\cite{Altarelli:2010gt, Ishimori:2010au, Kobayashi:2022moq, Ishimori:2012zz}. 
{\it But, research groups in refs. ~\cite{Heckman:2024obe,Kaidi:2024wio} have pointed out that the selection rules are dynamically broken at quantum levels even though they are invariant at the tree level.~\footnote{Systematic mechanisms for radiative corrections has been achieved by ref.~\cite{Kobayashi:2025cwx}.} 
This mechanism would not be realized by the other group-symmetries, and its feature would originate from a fact that the symmetry does not satisfy nature of group theory.}
Although such a nature sounds exotic, it is known that these kinds of symmetries appear in, {\it e.g.}, the  worldsheet theory  of perturbative string theory \cite{Bhardwaj:2017xup}, heterotic string theory on toroidal orbifolds \cite{Dijkgraaf:1987vp,Kobayashi:2004ya, Kobayashi:2006wq, Beye:2014nxa, Thorngren:2021yso, Heckman:2024obe, Kaidi:2024wio}, Calabi-Yau threefolds \cite{Dong:2025pah}, and type II intersecting/magnetized D-brane models \cite{Kobayashi:2024yqq,Funakoshi:2024uvy}.
}
{From a point of view of the building of phenomenological models, it is interesting to use the nature of these symmetries. For example, we can forbid some terms at tree level while they would be induced at loop level, and we would naturally explain a smallness of some quantities such as charged lepton and neutrino masses. }

In this work, we apply $\mathbb{Z}_3$ Tambara-Yamagami (TY) fusion rule to the lepton sector in a framework of lepton seesaw model~\cite{Nomura:2024ctl, Nomura:2025bph, Lee:2021gnw}, where TY is one of the non-invertible selection rules~\cite{Chang:2018iay}.
We realize a mass matrix component at one-loop level that should be smaller than the other components in the charged-lepton sector,
where a component of mass matrix is forbidden by the TY symmetry at tree level. On the other hand the neutrino mass is induced at one-loop level where the TY plays a role in an Abelian group such as $Z_2$ symmetry.
TY rule is simply given by
\begin{align}
\label{eq:TY}
    \eta^3 = \mathbb{I},\quad
    \eta \otimes {\cal N} = {\cal N},\quad
    {\cal N}\otimes {\cal N} = \mathbb{I} \oplus \eta \oplus \eta^2,
\end{align}
where $\{\eta, \eta^2,{\cal N}\}\otimes \mathbb{I}=\mathbb{I}\otimes \{\eta, \eta^2,{\cal N}\}=\{\eta, \eta^2,{\cal N}\}$.
Since the heavier Majorana neutrinos $N_R$ simultaneously contribute to the masses for charged-leptons and neutrinos,
we need three families of $N_R$. Otherwise, the charged-lepton masses would be vanished.
In addition, if we have TY symmetry only, there is no way to forbid  the SM charged-lepton Yukawa coupling ${\overline L_L H \ell_R}$ at the one-loop level,
which breaks the lepton seesaw mechanism.
Thus, we introduce the gauged $U(1)_{B-L}$ symmetry.
Thanks to this $B-L$ symmetry, not only the lepton seesaw can be retained, but also it assures three families of $N_R$ through the chiral anomaly cancellations.
Moreover, the symmetry provides a dominant annihilation cross section in order to explain the correct relic density of dark matter (DM) via Higgs portal interaction.
At the end, we show some numerical analysis including neutrino oscillation, charged-leptons, lepton flavor violations (LFVs), muon anomalous magnetic moment (muon $g-2$), and relic density of DM. 

This paper is organized as follows.
In Sec. \ref{sec:II}, 
we review our lepton seesaw model and explain how to induce a small masses of the charged-lepton masses as well as the neutrino masses.
Then, we discuss our DM candidate and show the relevant annihilation cross sections to explain the relic density.
Then, in sec. \ref{sec:III}, we demonstrate several numerical results satisfying neutrino observables, LFVs, muon $g-2$, relic density of DM in cases of normal hierarchy and inverted one.
In Sec. \ref{sec:IV}, we  summarize and conclude.

\section{Model setup}
\label{sec:II}

\begin{table}[t!]
\begin{tabular}{|c||c|c|c|c|c||c|c|c|c|c|}\hline\hline  
& ~$L_L$~ & ~$\ell_R$~ & ~${E_R}$~ & ~${E_L}$~ & ~${N_R}$~ & ~$H$~ & ~{$\eta$}~ & ~{$S^-$}~ & ~{$\varphi$}~ & ~{$\varphi$'}~  \\\hline\hline 
$SU(2)_L$   & $\bm{2}$  & $\bm{1}$  & $\bm{1}$ & $\bm{1}$ & $\bm{1}$ & $\bm{2}$  & $\bm{2}$    & $\bm{1}$  & $\bm{1}$  & $\bm{1}$  \\\hline 
$U(1)_Y$    & $-\frac12$  & $-1$  & $-1$ & $-1$  & $0$ & $\frac12$  & $\frac12$ & $-1$ & $0$& $0$    \\\hline
$TY$   & $\bm{1}$  & $ \bm \zeta $& $ \bm \zeta $ & $\bm \zeta$ & $\bm N $  & $\bm{1}$ & $\bm{N}$ & $\bm{N}$  & $\bm{1}$ & $\bm{1}$         \\\hline 
$U(1)_{B-L}$    & $-1$  & $-1$& $-3$ & $-3$ & $-1$ & $0$ & $0$  & $-4$  & $+2$  & $+4$     \\\hline
\end{tabular}
\caption{Charge assignments of the fermions and bosons
under $SU(2)_L\otimes U(1)_Y \otimes {TY} \otimes {U(1)_{B-L}}$, where all fermions have to have three generations.}\label{tab:1}
\end{table}
We summarize our model in this section.
We introduce three families of vector-like singly-charged fermions $E_{L,R}$, the three heavy neutral Majorana fermions $N_{R}$,
an inert scalar doublet $\eta\equiv [\eta^+,(\eta_R+i\eta_I)/\sqrt2]^T$, a singly-charged scalar $S^-$, and singlet scalars $\{\varphi,\ \varphi'\}$ in addition to the SM particles.
{\it In order to forbid the charged-lepton Yukawa interactions ${\overline L_L}H\ell_R$ at one-loop level after violating the TY fusion rule, 
we need to forbid the term $\overline{N_{R_a}^C} \ell_{R_b} S^+$. In fact, we have no discrimination between $\ell_R$ and $E_R$ due to the same assignments for the TY fusion rule. 
Thus, we introduce a gauged $U(1)_{B-L}$.
We assign $B-L$ charge of $E_{L,R}$ to be $-3$, $N_R$ to be $-1$, $\eta$ to be 0, $S^-$ to be $-4$,  $\varphi$ to be 2 and $\varphi'$ to be 4, where the charge for the SM lepton particles are assigned to be $-1$ as usual. 
}
The SM Higgs field is denoted by $H\equiv   [w^+,(v+h+iz)/\sqrt2]^T$, where Nambu-Goldstone bosons $w^+$ and $z$ are respectively eaten by the SM gauge bosons; $W^+$ and $Z$. The vacuum expectation value (VEV) of Higgs field is $v\approx 246$ GeV. The singlet scalars $\varphi$  and $\varphi'$ have nonzero VEVs which are denoted by $v_\varphi,\ v'_\varphi$, respectively, which break the gauged $B-L$ symmetry.
Then we impose the TY selection rule to these fields.
All their assignments are listed in Table \ref{tab:1}.
Under these symmetries, the valid Lagrangian is given by  
\begin{align}
-{\cal L}_\ell =
f_{ia} \overline{L_{L_i}} {\tilde \eta} N_{R_a} 
+
g_{ab} \overline{N_{R_a}^C} E_{R_b} S^+ 
+
(y_{E_2})_{ai} \varphi^* \overline{E_{L_a}}\ell_{R_i} 
+
(M_E)_{aa}  \overline{E_{L_a}} E_{R_a}
+
(y_N)_{aa} \varphi  \overline{N^C_{R_a}} N_{R_a}
+{\rm h.c.},
\label{eq:lpy}
\end{align}
where $\tilde\eta\equiv i\tau_2 \eta^*$ being $\tau_2$ the second Pauli matrix and we define $m_{E_2}\equiv y_{E_2} v_\varphi/\sqrt2$.
$M_E$ and $M_N\equiv y_N v_\varphi/\sqrt2$ can be diagonal without loss of generality. 

The valid Higgs potential is given by
\begin{align}
{\cal V} &=
-\mu_H^2 |H|^2 - \mu_\eta^2|\eta|^2 - \mu_S^2|S^-|^2 - \mu_\varphi^2|\varphi|^2 - \mu_{\varphi'}^2|\varphi'|^2
+\mu_\varphi (\varphi^{*2} \varphi' +{\rm c.c.})
+\lambda_0 \left(H^T(i\tau_2) \eta S^- \varphi' +{\rm h.c.}\right)
 \nn\\
&+\lambda_H |H|^4 +\lambda_\eta |\eta|^2 + \lambda_S  |S^-|^4-  \lambda_\varphi^2|\varphi|^4 -  \lambda_{\varphi'}^2|\varphi'|^4
+ \lambda_{H\eta} |H|^2|\eta|^2
+ \lambda'_{H\eta} |H^\dag\eta|^2
+ \lambda''_{H\eta} \left((H^\dag\eta)^2+{\rm h.c.}\right)\nn\\
&+ \lambda_{H S} |H|^2 |S^-|^2+ \lambda_{H \varphi} |H|^2 |\varphi|^2+ \lambda_{H \varphi'} |H|^2 |\varphi'|^2
+ \lambda_{\eta S} |\eta|^2 |S^-|^2 + \lambda_{\eta \varphi} |\eta|^2 |\varphi|^2+ \lambda_{\eta \varphi'} |\eta|^2 |\varphi'|^2
+ \lambda_{\varphi \varphi'} |\varphi|^2 |\varphi'|^2 .
\label{eq:lpy}
\end{align}
Especially, $\mu^2 \equiv \lambda_0v v'_\varphi/2$ contributes to the radiative charged-lepton mass matrix and $\lambda''_{H\eta}$ plays a role in generating the neutrino mass matrix at one-loop level as can be seen later.

\subsection{Active neutrino mass matrix}
The mass matrix of active neutrinos is simply indued at one-loop level~\cite{Ma:2006km} and the relevant Lagrangian is as follows:
\[
f_{ia} \overline{L_{L_i}} {\tilde \eta} N_{R_a} 
+
(M_N)_{aa}  \overline{N^C_{R_a}} N_{R_a}
- \lambda''_{H\eta} \left((H^\dag\eta)^2+{\rm h.c.}\right)
 +{\rm h.c.}
\]
Then, the neutrino mass matrix is given by
\begin{align}
(m_\nu)_{ij} = \frac{f_{ia} M_{N_a} f^T_{aj}}{(4\pi)^2}
\left[
\frac{m_R^2}{m_R^2-M_{N_a}^2} \ln\left[\frac{m_R^2}{M_{N_a}^2}\right]
-
\frac{m_I^2}{m_I^2-M_{N_a}^2} \ln\left[\frac{m_I^2}{M_{N_a}^2}\right]
\right],
\label{eq:neutmass}
\end{align}
where $m_R$ is the mass of $\eta_R$ and $m_I$ is that of $\eta_I$.
Here, we simply rewrite $m_\nu$ as $f F_{N} f^T$ where $F_N$ consists of $M_N$ multiplied by the loop function.
Suppose $m_\nu$ is diagonalized by $D_\nu = U^T_\nu m_\nu U_\nu$,
$f$ is then rewritten as follows~\cite{Casas:2001sr}:
\begin{align}
f= U^*_\nu D_\nu^{1/2} O_N F^{-1/2}_N\lesssim
 {4\pi},
\end{align}
where $O_N$ is three by three complex orthogonal matrix; $O_N^T O_N =O_N O_N^T= 1$ with three complex values.
The observed mixing matrix $U\equiv V_{e_L}^\dag U_\nu$, where $V_{e_L}$ is the mixing matrix to diagonalize the mass matrix for the charged-leptons
and we will define it in the next subsection.

The sum of neutrino mass $\sum D_\nu$ is constrained by the minimal standard cosmological model with CMB data as $\sum D_{\nu}\le$ 120 meV~\cite{Planck:2018vyg}, which is defined by 
\begin{align}
\sum D_\nu=D_{\nu_1}  + D_{\nu_2}+ D_{\nu_3}.
\end{align}
{Note that stronger constraints are obtained by combining CMB data and Baryon Acoustic Ocsillation (BAO) data~\cite{DESI:2024mwx}. Here we apply the conservative limit by CMB only.}
The effective mass for neutrinoless double beta decay $ m_{ee} $ is given by 
\begin{align}
 m_{ee}= | D_{\nu_1} \cos^2\theta_{12} \cos^2\theta_{13}+ D_{\nu_2} \sin^2\theta_{12} \cos^2\theta_{13}e^{i\alpha_{21}}+ D_{\nu_3} \sin^2\theta_{13}e^{i(\alpha_{31}-2\delta_{CP})}|.
\end{align}
where $s_{12,23,13}$, which are short-hand notations $\sin\theta_{12,23,13}$, are neutrino mixing of $U$, $\delta_{CP}$ is Dirac phase, $\alpha_{21,31}$ are Majorana phases.
$ m_{ee} $ is constrained by the current KamLAND-Zen data measured in future~\cite{KamLAND-Zen:2024eml},
 and the current  upper bound is given by $ m_{ee} <(36-156)$ meV at 90 \% confidence level. \\
$ m_{\nu e} $ is also given by 
\begin{align}
 m_{\nu e}=\sqrt{D_{\nu_1}^2 c^2_{13} c^2_{12} + D_{\nu_2}^2 c^2_{13} s^2_{12}+D_{\nu_3}^2 s^2_{13}},
\end{align}
which is constrained by KATRIN~\cite{KATRIN:2024cdt}; $m_{\nu e}\le450$ meV at 90\% CL.

\begin{figure}[tb]\begin{center}
\includegraphics[width=100mm]{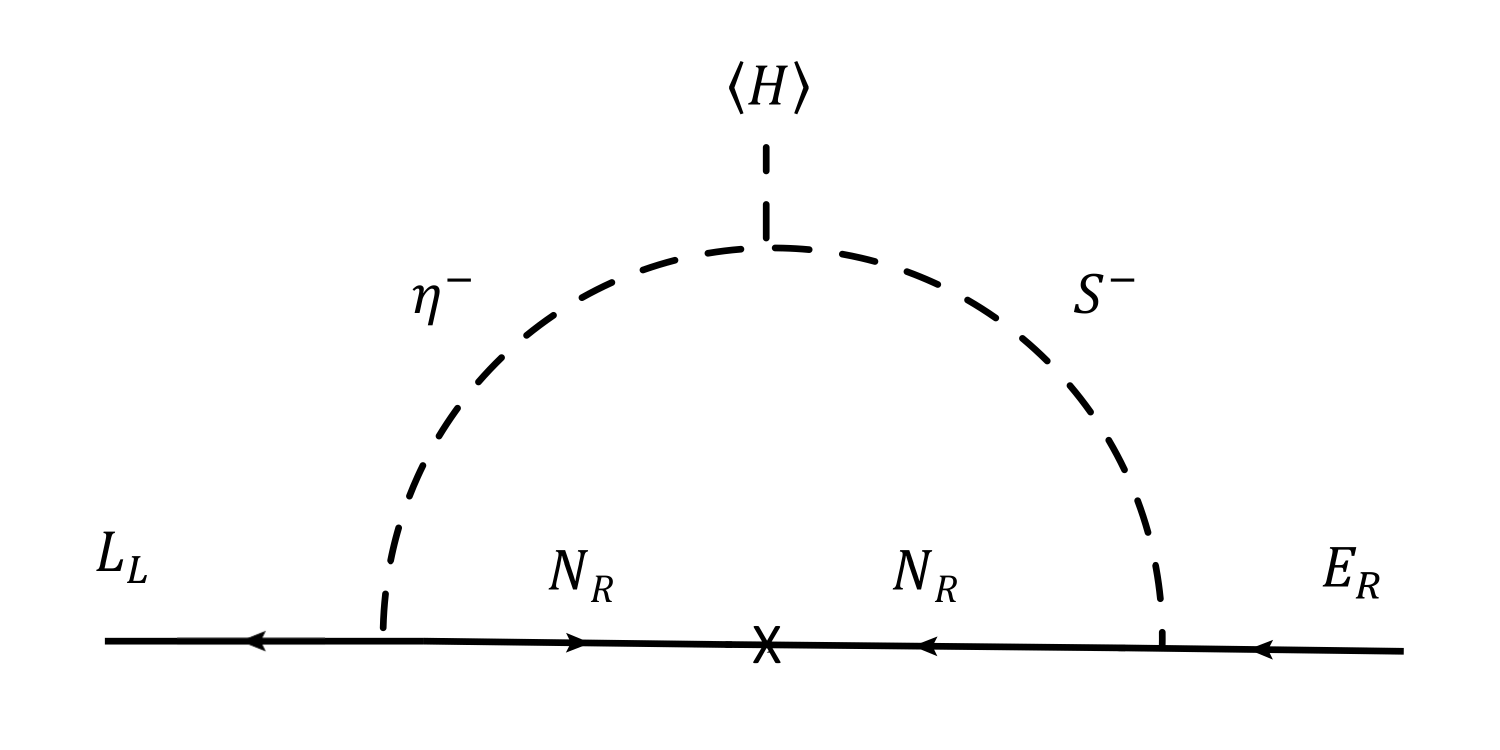}
\caption{A one-loop diagram that induces $\overline{L_L} H E_R$ operator.} 
\label{fig:diagram}
\end{center}\end{figure}

\subsection{Charged-lepton mass matrix}
The charged-lepton mass matrix at tree level is found by
\begin{align}
\left(
\begin{array}{c} 
\overline{\ell_L} \\ \overline{E_L} 
\end{array} \right)^T
  \left(\begin{array}{cc} 
0 & 0 \\
m_{E_2} &M_{E}
  \end{array} \right)
\left(\begin{array}{c} 
{\ell_R} \\ {E_R}
  \end{array} \right)
\label{eq:cgd-mtrx_tree}.
\end{align}
Clearly, the above mass matrix has three vanishing mass eigenvalues and all the masses of charged-leptons are zero.
Then, let us consider the terms 
\[
f_{ia} \overline{L_{L_i}} {\tilde \eta} N_{R_a} 
+
g_{ab} \overline{N_{R_a}^C} E_{R_b} S^+ 
-\mu^2 \eta^+ S^- +{\rm h.c.}\ .
\]
From these terms, one finds that there is the mass term of $(m_{E_1})_{ia} \overline{\ell_{L_i}} E_{R_a}$, which is originated from $\overline{L_{L}} \langle H \rangle E_{R}$, at one-loop level shown in Fig.~\ref{fig:diagram}.
The explicit computation of this mass matrix is given as
\begin{align}
(m_{E_1})_{ib} \approx \frac{\mu^2}{\sqrt2(4\pi)^2 \left({m_S^2}-{m_\eta^2} \right)} \sum_a 
\frac{f_{ia} M_{N_a}^{3} g_{a b}}
{\left({m_S^2}-M_{N_a}^2\right)}
\left[
\frac{m_S^2}{M_{N_a}^2}\ln\left[\frac{m_S^2}{M_{N_a}^2}\right]
-
\frac{m_\eta^2}{M_{N_a}^2}\ln\left[\frac{m_\eta^2}{M_{N_a}^2}\right]
\right],
\end{align}
where $m_\eta$ is the mass of $\eta^\pm$. In our convenience, we redefine $m_{E_1}\equiv f G_N g$. 
Finally, we find the correct  mass matrix:
\begin{align}
\left(
\begin{array}{c} 
\overline{\ell_L} \\ \overline{E_L} 
\end{array} \right)^T
  \left(\begin{array}{cc} 
0 & m_{E_1} \\
m_{E_2} &M_{E}
  \end{array} \right)
\left(\begin{array}{c} 
{\ell_R} \\ {E_R}
  \end{array} \right)
\label{eq:cgd-mtrx}.
\end{align}
Thus, we obtain the following mass hierarchy 
\begin{align}
m_{E_1}, \ m_{E_2}\ll M_E ,
\end{align}
where we assume $m_{E_2},\ll M_E$ by tuning the model parameters.~\footnote{We have a solution to forbid $m_{E_2}$ at tree level but allowed at one-loop level. However, the one-loop mass masses includes logarithmic divergence. Thus, we would not be able to adopt such a notation.}
Under the hierarchy, the mass matrix is approximately block-diagonalized by ${\rm diag}[m_{\ell}, M_{\ell}] \approx V_L^\dag {\cal M}_{\ell'} V_R$,
where 
\begin{align}
& V_L \simeq
\left(\begin{array}{cc} 
{\bf1} &  \theta_L \\
- \theta_L & {\bf 1}
  \end{array} \right),\quad
 V_R \simeq
\left(\begin{array}{cc} 
{\bf1} &  \theta_R \\
- \theta_R & {\bf 1}
  \end{array} \right), \\
& m_{\ell} \approx - m_{E_1} M_E^{-1} m_{E_2} ,\quad M_{\ell} \approx M_E,\\
& \theta_L\approx (M_E^\dag)^{-1} m^\dag_{E_1},\quad 
\theta_R\approx (M_E^\dag)^{-1} m^\dag_{E_2},
\end{align}
where  $M_E$ is diagonal.
The charged lepton mass matrix $m_{\ell}$ is further diagonalized by $D_\ell =V_{e_L}^\dag m_{\ell} V_{e_R}$. 
Therefore, we obtain $|D_\ell |^2=V^\dag_{e_L} m_{\ell} m_{\ell}^\dag V_{e_L}$ 
where $D_\ell=$diag.$(m_e,m_\mu,m_\tau)$.\\
%
Furthermore, we can rewrite $m_{E_1}$ and $m_{E_2}$ in terms of $V_{e_L}$, $D_\ell$, and some parameters as follows:
\begin{align}
m_{E_1}=i V_{e_L} D_\ell^{1/2} V_E^\dag M_E^{1/2},\quad
m_{E_2}=i M_E^{1/2} V_E D_\ell^{1/2} V_{e_R}^\dag,
\end{align}
where $V_E$ is an arbitrary unitary matrix $V_E^\dag V_E=1$. 
Applying the degrees of freedom of $V_E$ for $m_{E_2}$, we work on the diagonal basis of $m_{E_2}$.
Reminding $m_{E_1}=f G_N g$ and $V_{e_L}= U_\nu U^\dagger$, we rewrite $g$ as follows:
\begin{align}
g = i G_N^{-1} f^{-1} V_{e_L} D_\ell^{1/2} V_E^\dag \sqrt{M_E}
= i G_N^{-1} 
F_N^{1/2} O_N^{-1} D_\nu^{-1/2} U_\nu^T U_\nu U^\dag
 D_\ell^{1/2} V_E^\dag \sqrt{M_E}
 \lesssim {4\pi}.
\end{align}

\subsection{Lepton flavor violations and muon $g-2$}
 The dominant LFVs and muon $g-2$ are induced via the following terms:
 \begin{align}
 F_{\sigma a} \overline{\ell_{L_\sigma}} N_{R_a}\eta^- -G_{a\rho}\overline{N^C_{R_a}} \ell_{R_\rho} S^+ -\mu^2 \eta^+ S^-+{\rm h.c.},
\end{align}
where $F\equiv V^\dag_{e_L} f$, $G\equiv g\theta_R$, and all the above particles are supposed to be mass eigenstates.
Then, the branching ratios for LFVs and muon $g-2$ are given by
  \begin{align}
&{\rm BR}(\ell_a\to \ell_b\gamma)\approx \frac{3\mu^4}{64\pi} 
\frac{\alpha_{em}C_{ab}}{G^2_F m_{\ell_a}^2}
\left[
\left| F^*_{a\alpha} (M'^{-3})_\alpha G^*_{\alpha b} \right|^2
+
\left| G^T_{a\alpha} (M'^{-3})_\alpha F^T_{\alpha b} \right|^2
\right], \\
& \Delta a_\mu\approx  - \frac{\mu^2}{2(4\pi)^2} m_\mu
\left[ F^*_{2\alpha} (M'^{-3})_\alpha G^*_{\alpha 2} + G^T_{2\alpha} (M'^{-3})_\alpha F^T_{\alpha 2} \right],\\
& (M'^{-3})_\alpha \approx \frac1{M_{N_\alpha}^3} 
\int[dx]_4(2-b-c-d)
 \left[ \frac{1}{[a + b r_{S\alpha} +(c+d) r_{\eta\alpha} ]^2}  + \frac{1}{[a + (b+d) r_{S\alpha} +c r_{S\eta}]^2} \right],
\end{align}
where $ r_{S\alpha}\equiv m_S^2/M_{N_\alpha}^2$, $ r_{\eta\alpha}\equiv m_\eta^2/M_{N_\alpha}^2$, $[dx]_4\equiv (da)(db)(dc)(dd)\delta(1-a-b-c-d)$,
$G_F$ is Fermi constant, $\alpha_{em}$ is fine structure constant, and $C_{21}\approx1$, $C_{31}\approx0.1784$, $C_{32}\approx0.1736$. 
Experimental upper bounds for the branching ratios of the LFV decays are given by~\cite{MEG:2016leq, BaBar:2009hkt,Renga:2018fpd,MEGII:2023ltw}
\begin{align}
{\rm BR}(\ell_\mu\to \ell_e\gamma)\lesssim 3.1\times10^{-13},\quad
{\rm BR}(\ell_\tau\to \ell_e\gamma)\lesssim 3.3\times10^{-8},\quad
{\rm BR}(\ell_\tau\to \ell_\mu\gamma)\lesssim 4.4\times10^{-8}.
\end{align}
Muon $g-2$ is recently close to the prediction of the SM~\cite{Aliberti:2025beg} adopting current Lattice results which is within 1$\sigma$;
\begin{align}
\Delta a_\mu \simeq (39\pm 64)\times 10^{-11}.
\end{align}
It implies that negative value is also allowed now.

\subsection{Dark matter candidate}
Here, we assume the lightest neutral fermion to be the DM candidate, where it is denoted by $(N_{R_1}\equiv)X_R$ and its mass is $m_\chi$.
The dominant annihilation cross section to explain the relic density of the DM arises from the same terms as LFVs and muon $g-2$:
 \begin{align}
 F_{\sigma 1} \overline{\ell_{L_\sigma}} X_{R}\eta^- -G_{1\rho}\overline{X^C_{R}} \ell_{R_\rho} S^+ -\mu^2 \eta^+ S^-+{\rm h.c.}.
\end{align}
The cross section $\sigma v_{rel}$ is expanded by the relative velocity $v_{rel}$ as $\sigma v_{rel}\approx a + b v_{rel}^2 +{\cal O}(v_{rel}^4)$,
and $a$ and $b$ are respectively given by
\begin{align}
a&\approx  \sum_{\sigma,\rho=1}^3\frac{|F_{\sigma 1} G_{1\rho}|^2}{16\pi} \frac{\mu^4 m_\chi^2}{(m_\chi^2+m_S^2)^2(m_\chi^2+m_\eta^2)^2},\label{eq.s}\\
b&\approx \sum_{\sigma,\rho=1}^3 \frac{|F_{\sigma 1} G_{1\rho}|^2}{96\pi} \frac{\mu^4 m_\chi^2}{(m_\chi^2+m_S^2)^4(m_\chi^2+m_\eta^2)^4}\nn\\
&\times
\left[
 3 m_\chi^8 -4 m_\chi^6 m_\eta^2 -m_\chi^4 m_\eta^4 - m_S^4(m_\chi^4+4 m_\chi^2 m_\eta^2-3m_\eta^4) -4m_\chi^2 m_S^2
 (m_\chi^4 + 5 m_\chi^2 m_\eta^2+ m_\eta^4)
\right],\label{eq.p}
\end{align}
 where $v_{rel}^2\approx 0.2$.
 The typical range of the cross section to describe the correct relic density at 2$\sigma$; $0.1196\pm2\times0.0031$~\cite{Planck:2013pxb} is
 \begin{align}
 1.77552\le (\sigma v_{rel}) \times 10^9\ {\rm GeV}^2 \le 1.96967.
 \end{align}
 However since our cross section in Eqs.~(\ref{eq.s}) and (\ref{eq.p}) is 10$^{-14}$ GeV$^{-2}$ at most for both the cases of NH and IH in our numerical analysis,
 we rely on a resonance via Higgs portal cross section that is p-wave dominant~\cite{Kajiyama:2013zla}; 
 \begin{align}
b&\approx 
\frac{m_\chi^2 s^2_\alpha c^2_\alpha}{8\pi v^2 v_\varphi^2}
\left|\frac1{4 m_\chi^2 -m_h^2 +i m_h\Gamma_h }\right|^2
\sum_{i=f,V}
\sqrt{1-\frac{m_i^2}{m_\chi^2}}\nn\\
&\times 
\left[
3 m_i^2 m_\chi^2 \left(1-\frac{m_i^2}{m_\chi^2}\right)\delta_{if}
+
m_\chi^4  \left(\frac32-2\frac{m_\chi^2}{m_i^2} +2\frac{m_\chi^4}{m_i^4} \right)\delta_{iV} 
\right],
\label{eq.p-hp}
\end{align}
 where $f$ is all the standard model fermions, $s_\alpha$ is mixing between $h$ and $\varphi$, $V$ is gauge bosons $Z_0$ and $W^\pm$, and $\Gamma_h\equiv 3.2\times10^{-3}$ GeV is the total decay rate of the SM Higgs. 
 Note here that we neglect the second Higgs portal contribution. 
 We adopt experimental values for the SM masses in PDG~\cite{ParticleDataGroup:2022pth}.

 \section{Numerical results}
 \label{sec:III}
 Here, we show numerical analysis and display some useful figures.
 We apply for the best fit values of the neutrino oscillation data referred to Nufit 6.0~\cite{Esteban:2024eli} without SK atmospheric data.
Then, we randomly select our input parameters as follows:
\begin{align}
& 
10 \le \frac{M_{N_1}(\equiv m_\chi)}{\rm GeV} \le 10^3, \quad
1.2\times M_{N_1}  \le \frac{M_{N_2}}{\rm GeV}  \le 10^4, \quad
1.2\times M_{N_1}  \le \frac{M_{N_3}}{\rm GeV}   \le 10^4 , \\
&
10^{-3} \le \frac{\delta m}{\rm GeV}  \le 1 , \quad
10^{-3} \ {\rm GeV}\le \frac{\mu}{\rm GeV}  \le 10^3 , \quad
1.2\times M_{N_1} \le \frac{m_S}{\rm GeV}  \le 10^4 , \quad
1.2\times M_{N_3} \le \frac{m_I}{\rm GeV}  \le 10^5, \nn\\
&
1.2\times m_{E_1}  \le \frac{M_{E_1}}{\rm GeV}  \le 10^8, \quad
1.2\times M_{E_1}  \le \frac{M_{E_2}}{\rm GeV}   \le 10^8, \quad 
1.2\times M_{E_2}  \le \frac{M_{E_3}}{\rm GeV}   \le 10^8, \nn\\%
&
1.2\times M_{N_1}  \le \frac{(m_{E_2})_{11}}{\rm GeV}  \le 10^8, \quad
1.2\times M_{N_1}  \le \frac{(m_{E_2})_{22}}{\rm GeV}   \le 10^8, \quad 
1.2\times M_{N_1} \le \frac{(m_{E_2})_{33}}{\rm GeV}   \le 10^8, \quad \nn
\end{align}
where all the mixing angles such as Majorana phases, $O_N,\  V_E$, run whole  the range; $[-\pi,+\pi]$, $m_R=m_I+\delta m$, and $m_\eta\equiv m_I$  simply to satisfy the oblique parameters.
Under the above regions, we impose a perturbative limit for $G$ and $F$;
 \begin{align}
 {\rm Max}\left(|F|,\ |G|\right)\lesssim 4\pi.
\end{align}

\subsection{NH}

\begin{figure}[tb]\begin{center}
\includegraphics[width=53mm]{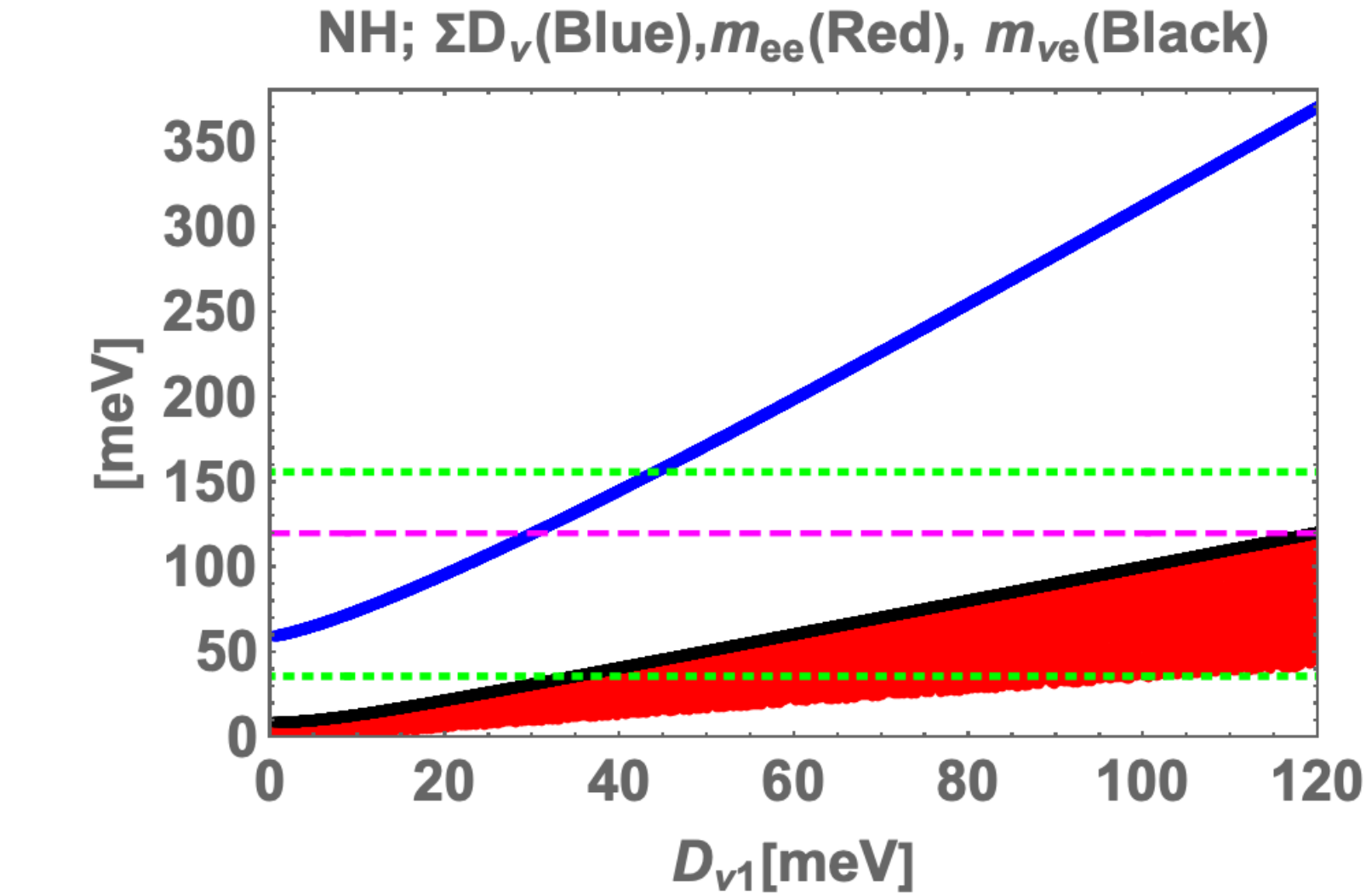}
\includegraphics[width=53mm]{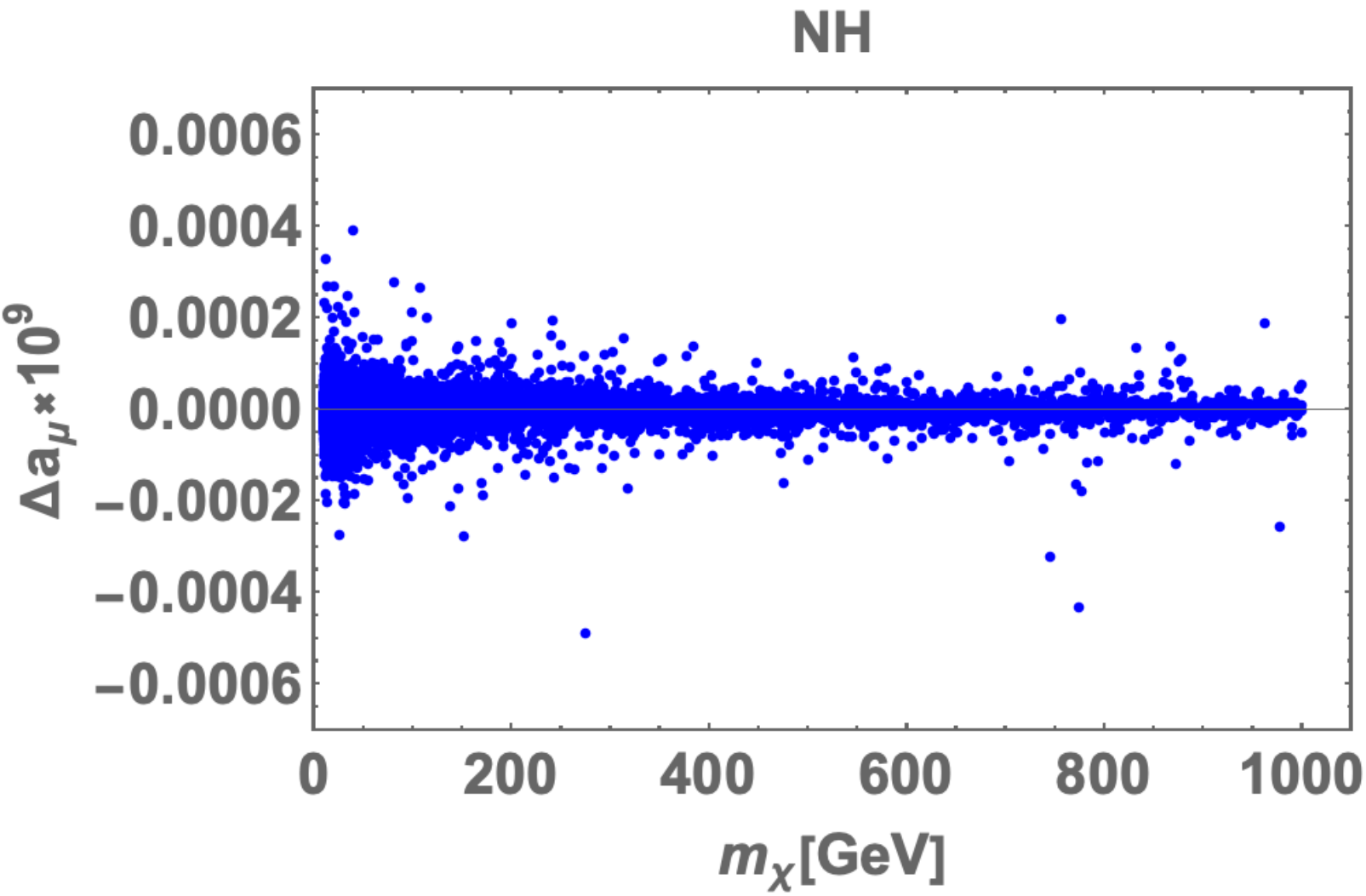}
\includegraphics[width=53mm]{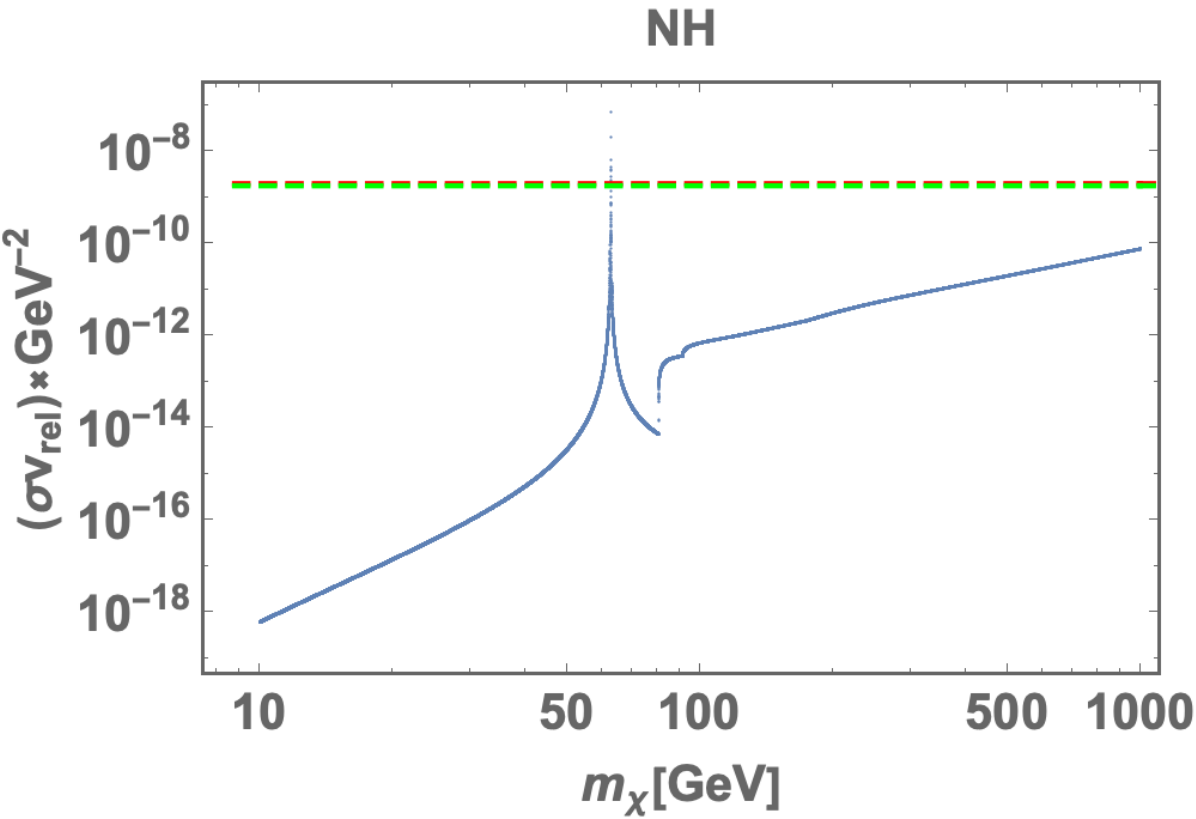}
\caption{Allowed regions for $\{\sum D_\nu,\ m_{ee},\ m_{\nu e}\}$ (left)  in terms of $D_{\nu_1}$ in meV unit, $\Delta a_\mu\times10^9$ (center), and $(\sigma v_{rel})\times {\rm GeV}^{-2}$ (right), in terms of DM mass $N_{R_1}$
in GeV unit. Here the magenta dashed horizontal line in the left figure is the upper limit of the minimal standard cosmological model $120$ meV, {and the green ones are the upper limits of $m_{ee}$ for the neutrinoless double beta decay; $m_{ee} \le 36(156)$ meV for tightest(loosest) bound.}
The green region is our allowed ones for $\sum D_\nu$, red one $m_{ee}$, and black one $m_{\nu e}$.
The green and red horizontal dashed liens in the right figure  represent upper and lower bound of cross section to explain the correct relic density of DM $\sim{\cal O}(10^{-9})$ GeV$^{-2}$.} 
\label{fig:nh1}
\end{center}\end{figure}
%
Fig.~\ref{fig:nh1} shows the values of $\{\sum D_\nu,\ m_{ee},\ m_{\nu e}\}$ (left) in terms of $D_{\nu_1}$ in meV unit, $\Delta a_\mu\times10^9$ (center), and $(\sigma v_{rel})\times {\rm GeV}^{-2}$ (right), in terms of DM mass $N_{R_1}$
in GeV unit. Here the magenta dashed horizontal line in the left figure is the upper limit of the minimal standard cosmological model $120$ meV, {and the green ones are the upper limits of $m_{ee}$ for the neutrinoless beta decay; $m_{ee} \le 36(156)$ meV for tightest(loosest) bound.}
The blue, red and green regions are our predictions for $\sum D_\nu$, $m_{ee}$, and $m_{\nu e}$ respectively in the left figure.
The green and red horizontal dashed liens in the right figure represent the upper and the lower bound of the annihilation cross section to explain the correct relic density of DM $\sim{\cal O}(10^{-9})$ GeV$^{-2}$.
The left figure tells us that $D_{\nu_1}\lesssim 30$ meV that
arises from the minimum standard cosmological model $\sum D_\nu\le120$ meV. 
The center figure suggests that $|\Delta a_\mu|$ is of the order $5\times10^{-13}$ at most.
The right figure shows that the correct relic density is satisfied at nearby the half mass of the SM Higgs $m_h/2$ GeV as is expected.

\begin{figure}[tb]\begin{center}
\includegraphics[width=53mm]{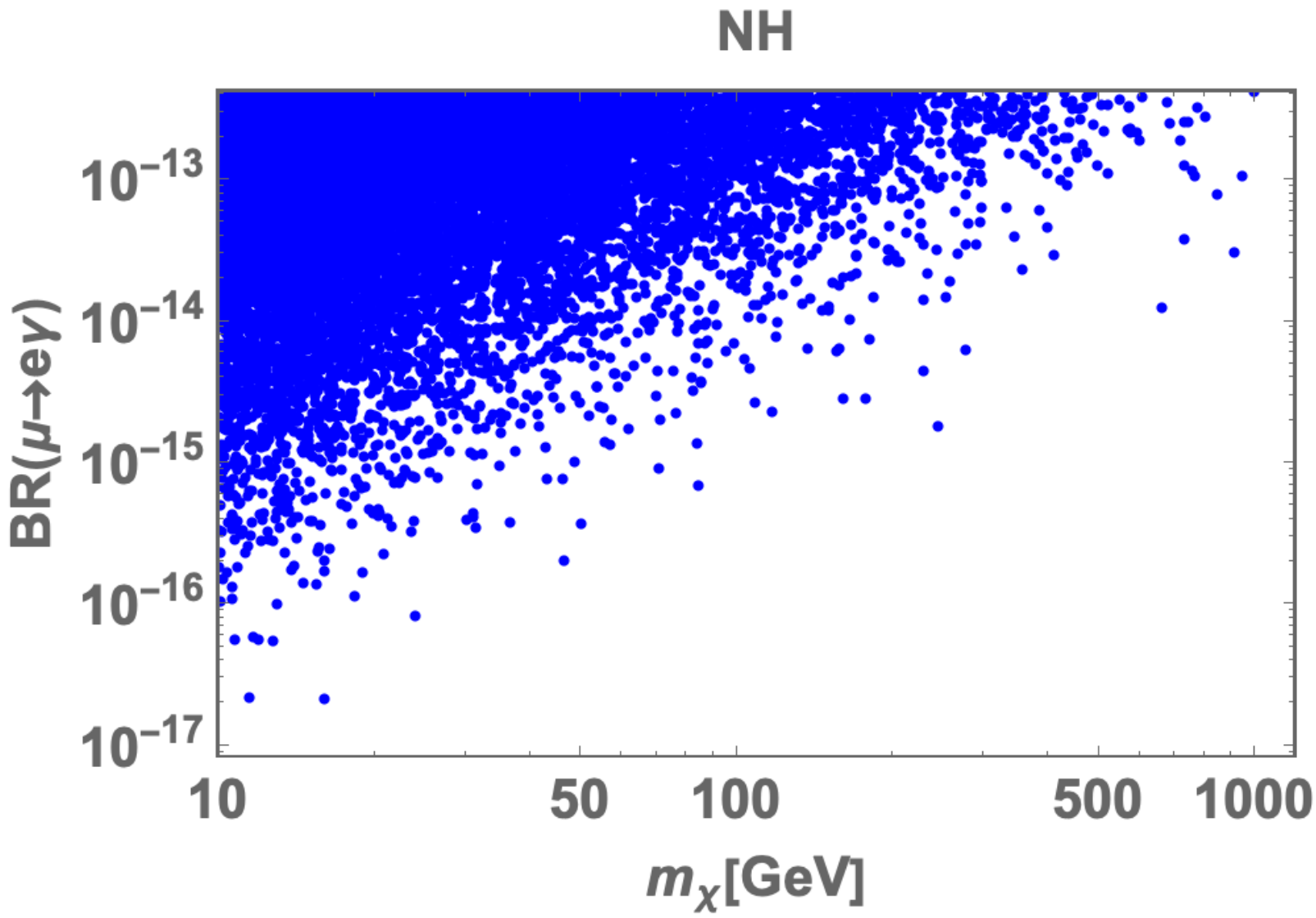}
\includegraphics[width=53mm]{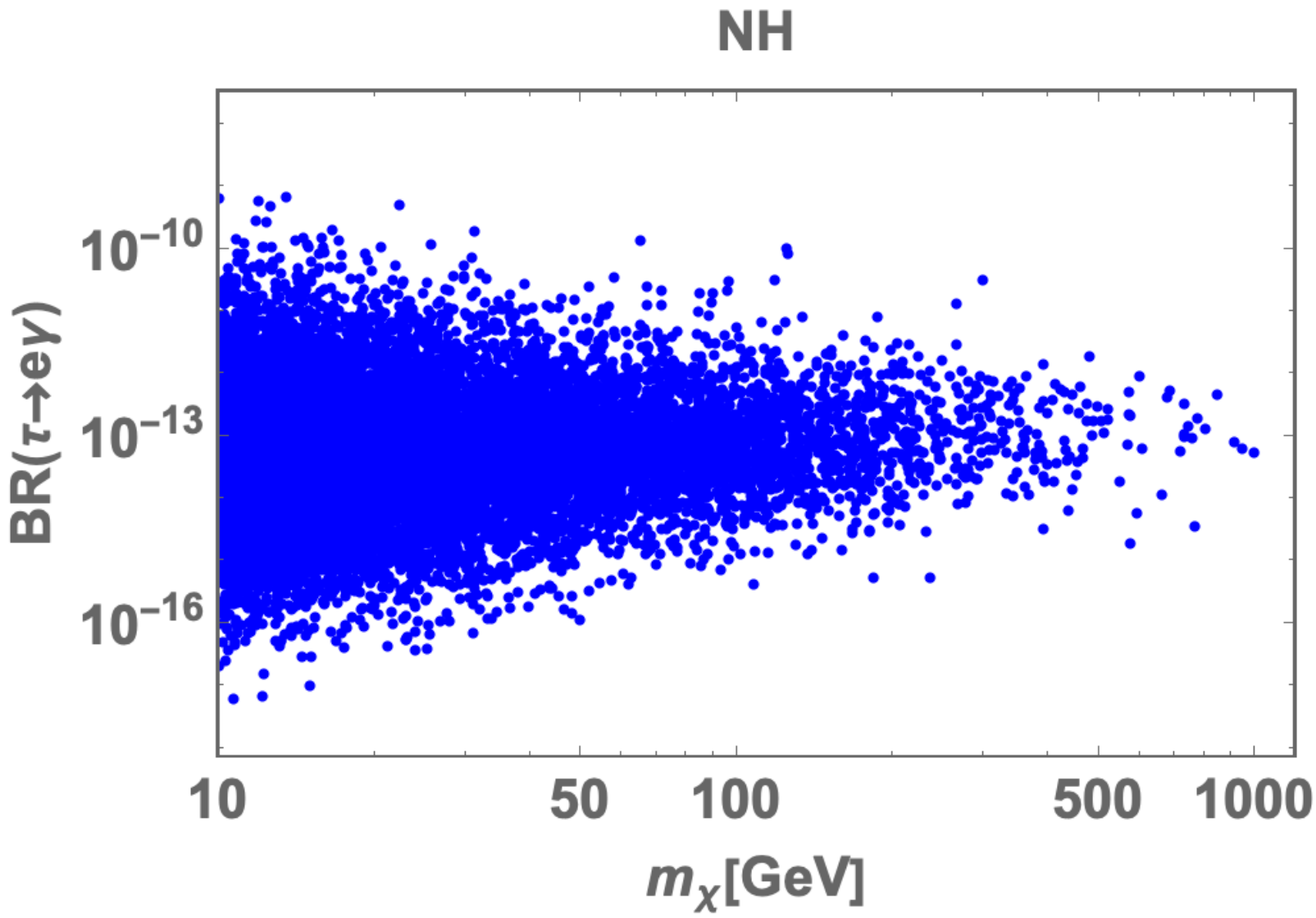}
\includegraphics[width=53mm]{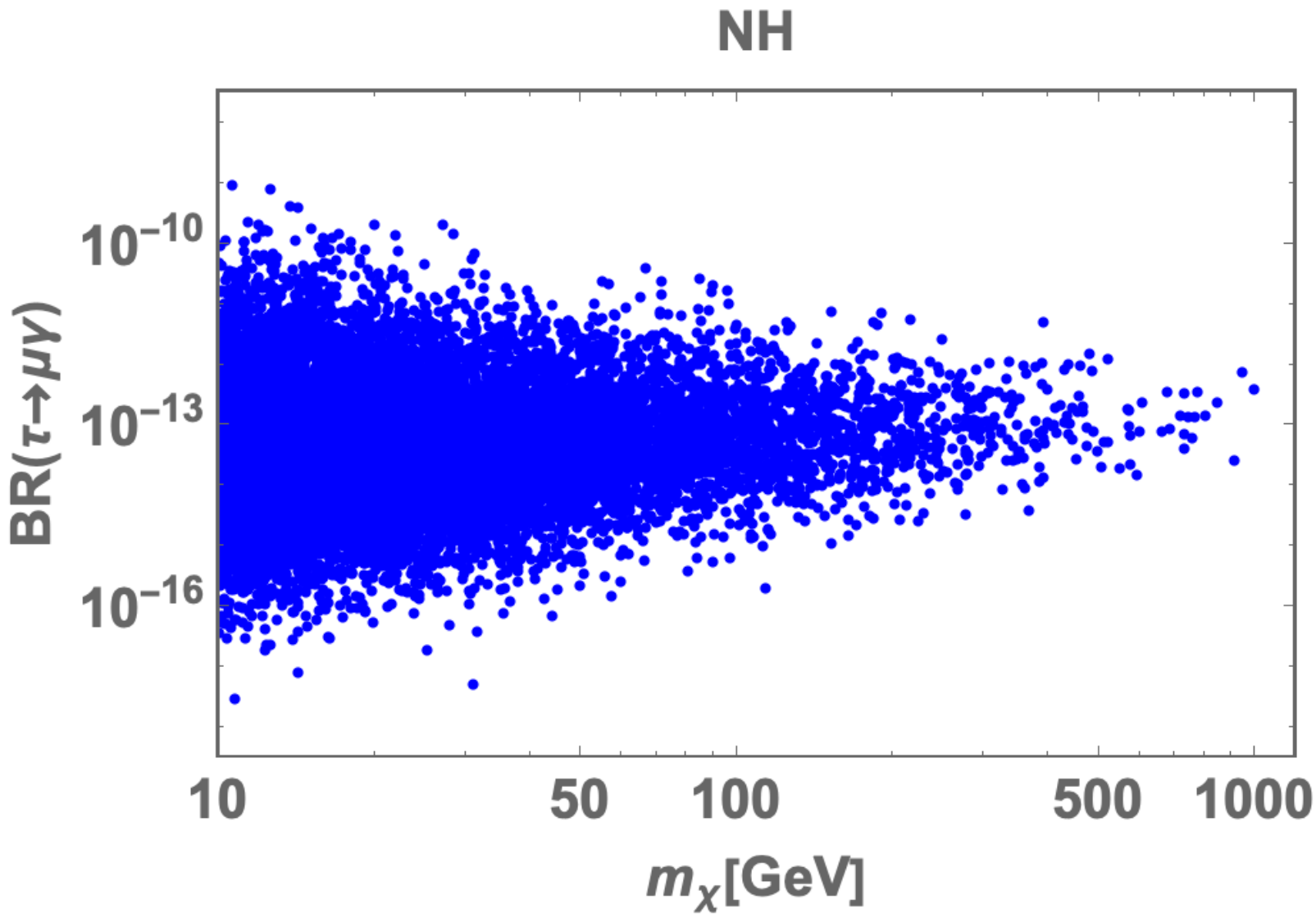}
\caption{Allowed regions for BR($\mu\to e\gamma$)(left), BR($\tau\to e\gamma$)(center), BR($\tau\to \mu\gamma$)(right) in terms of DM mass $N_{R_1}$.} 
\label{fig:nh2}
\end{center}\end{figure}
%
Fig.~\ref{fig:nh2} shows three LFVs; BR($\mu\to e\gamma$)(left), BR($\tau\to e\gamma$)(center), BR($\tau\to \mu\gamma$)(right) in terms of DM mass $N_{R_1}$.
These figures imply that the stringent constraint comes from  BR($\mu\to e\gamma$) that is well-tested near future.

\subsection{IH}

\begin{figure}[tb]\begin{center}
\includegraphics[width=53mm]{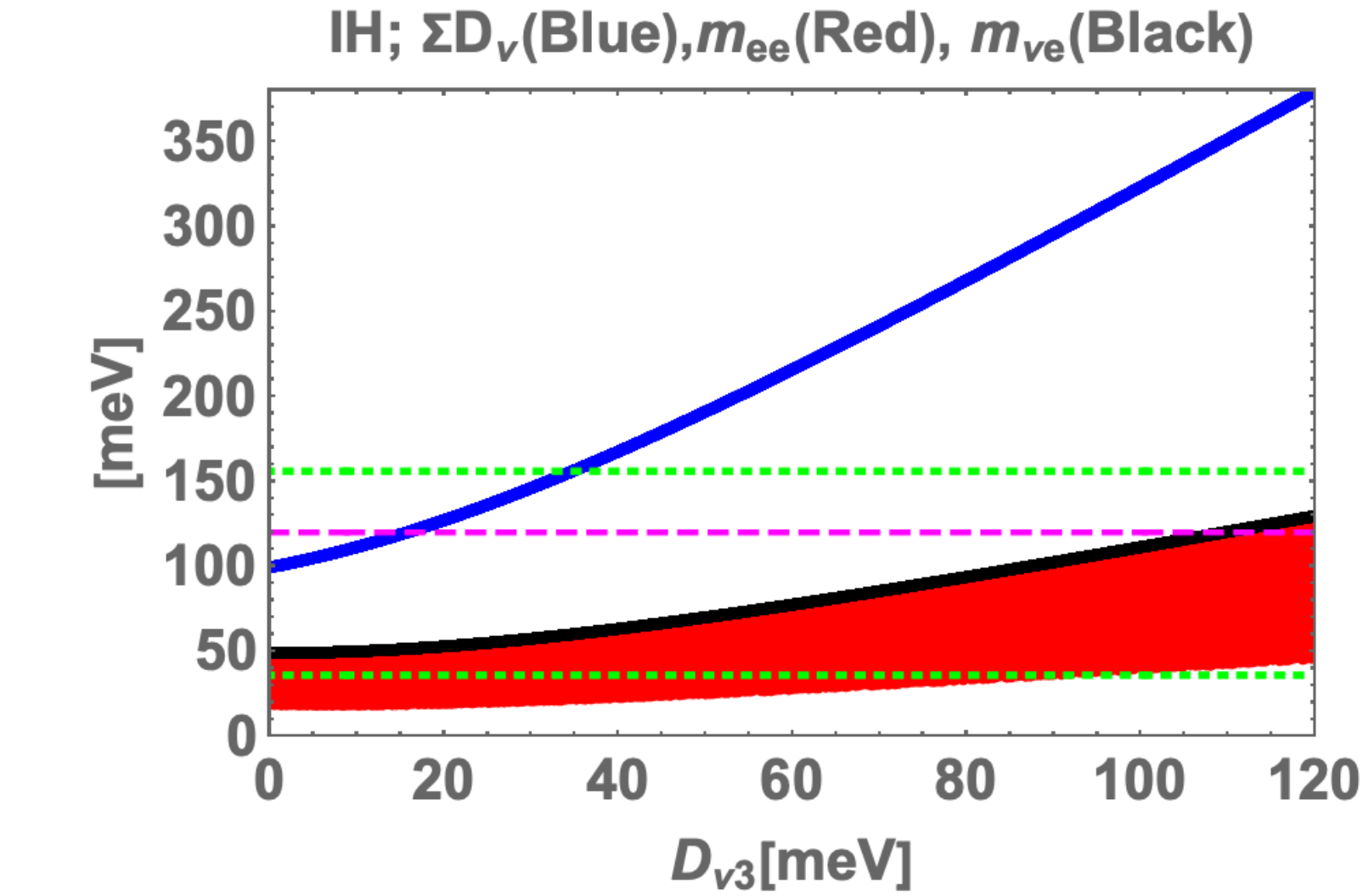}
\includegraphics[width=53mm]{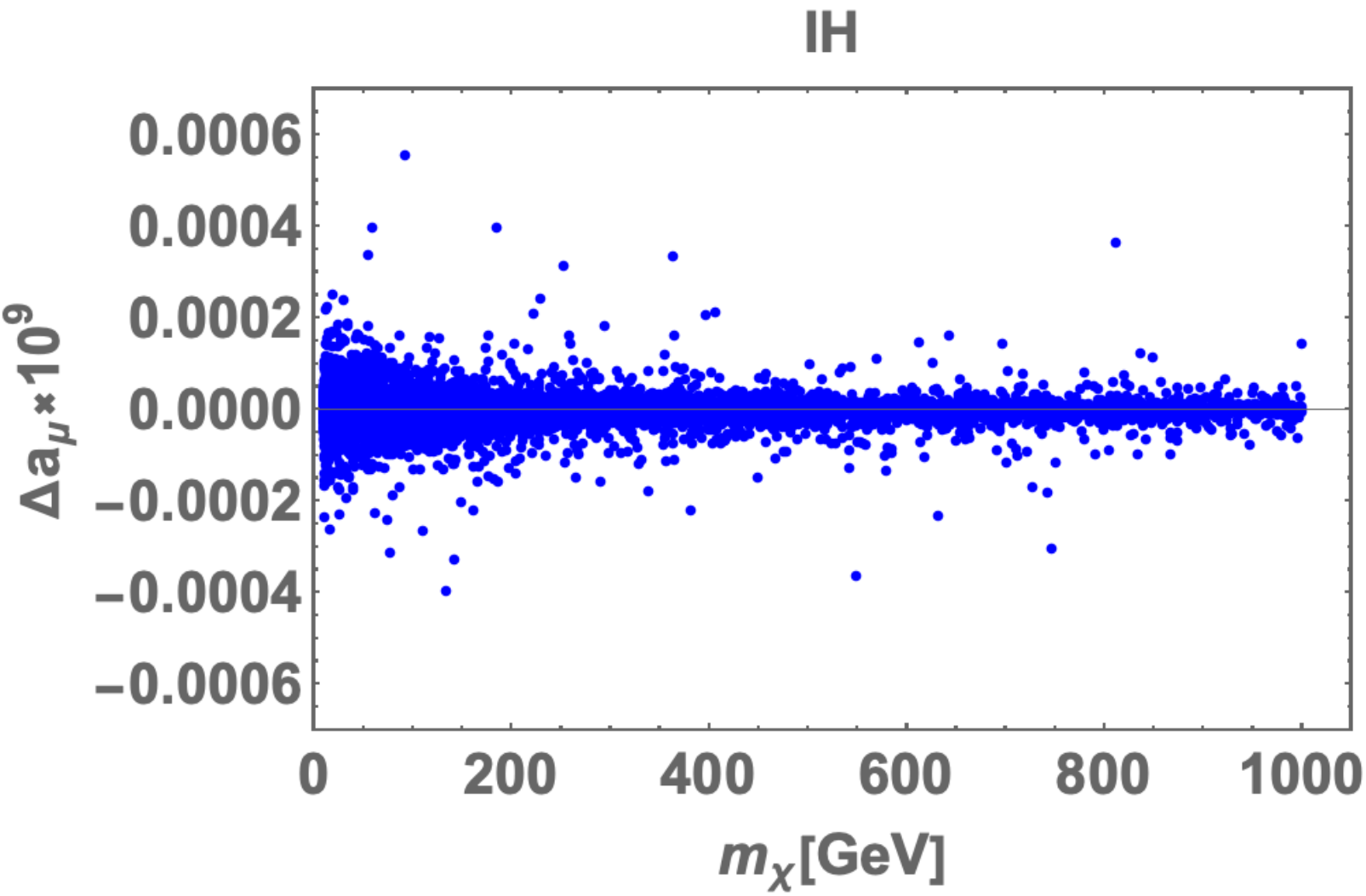}
\includegraphics[width=53mm]{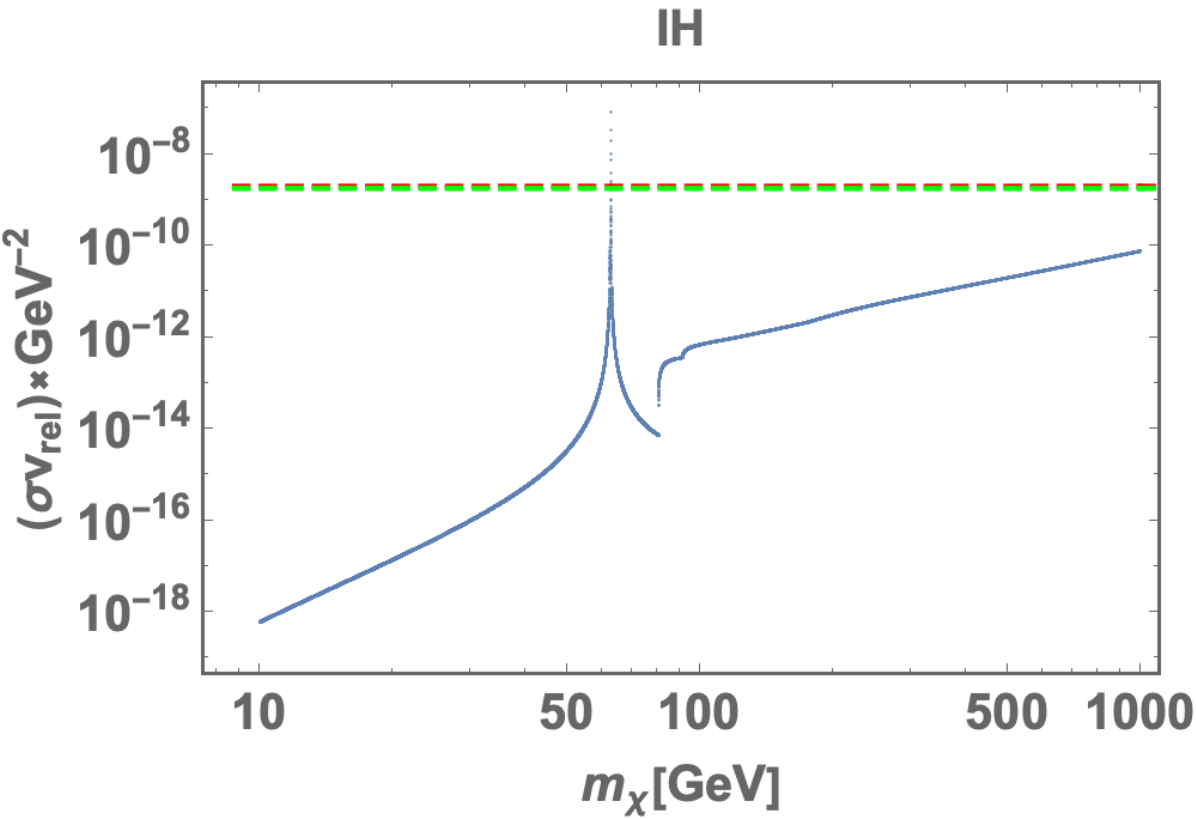}
\caption{Allowed regions for $\sum D_\nu,\ m_{ee},\ m_{\nu e}$(left)  in terms of $D_{\nu_1}$ in meV unit, $\Delta a_\mu\times10^9$ (center), and $(\sigma v_{rel})\times {\rm GeV}^{-2}$ (right), in terms of DM mass $N_{R_1}$
in GeV unit, where the legends are the same as Fig.~\ref{fig:nh1}. } 
\label{fig:ih1}
\end{center}\end{figure}
%
Fig.~\ref{fig:ih1} shows the values of $\{\sum D_\nu,\ m_{ee},\ m_{\nu e}\}$ (left) in terms of $D_{\nu_3}$ in meV unit, $\Delta a_\mu\times10^9$ (center), and $(\sigma v_{rel})\times {\rm GeV}^{-2}$ (right), in terms of DM mass $N_{R_1}$
in GeV unit, where the legends are the same as Fig.~\ref{fig:nh1}.
The left figure tells us that $D_{\nu_1}\lesssim 20$ meV that
arises from the minimum standard cosmological model $\sum D_\nu\le120$ meV. 
The center figure suggests that $|\Delta a_\mu|$ is of the order $6\times10^{-13}$ at most.
The right figure shows that the correct relic density is satisfied at nearby the half mass of the SM Higgs $m_h/2$ GeV as is expected.

\begin{figure}[tb]\begin{center}
\includegraphics[width=53mm]{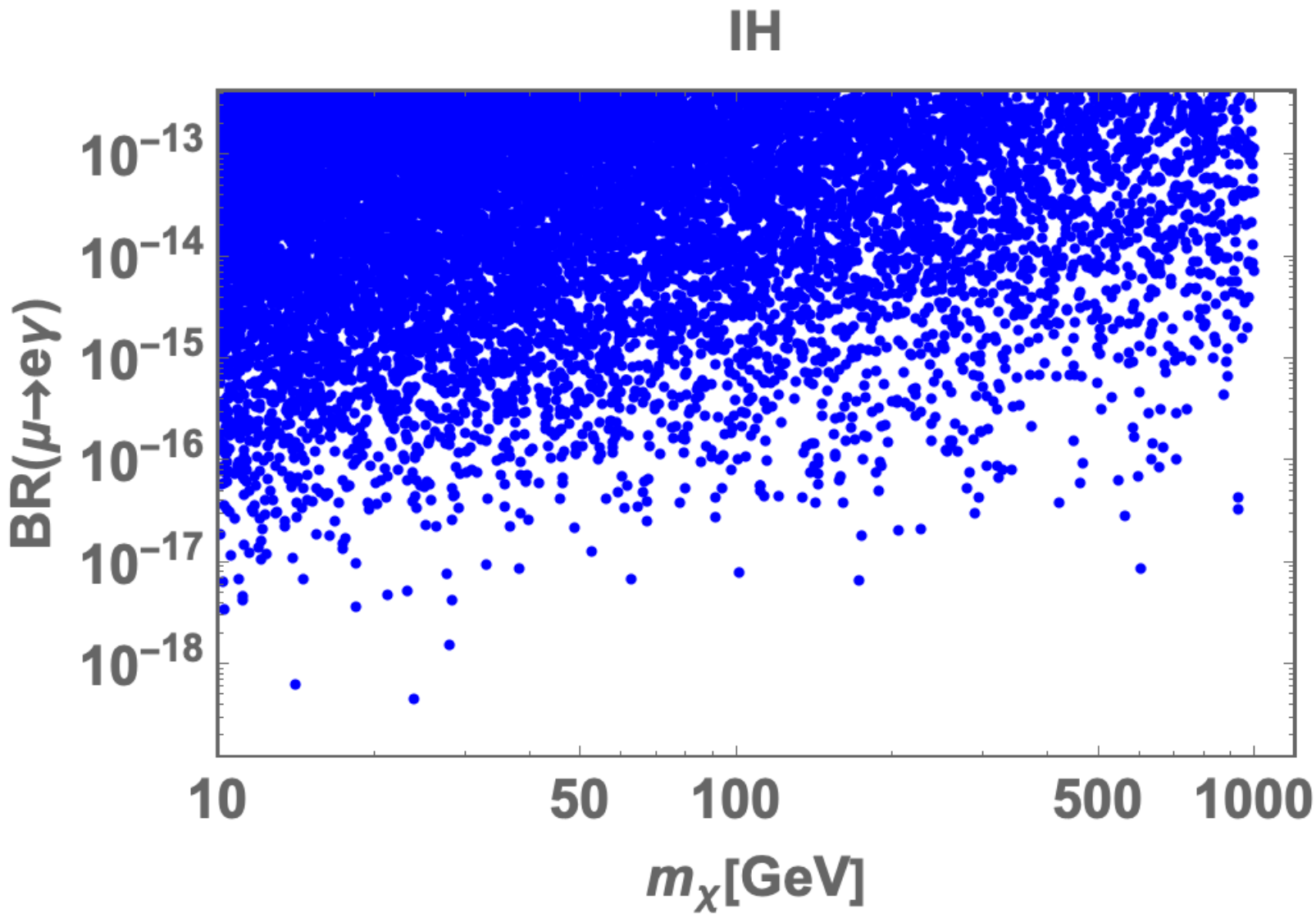}
\includegraphics[width=53mm]{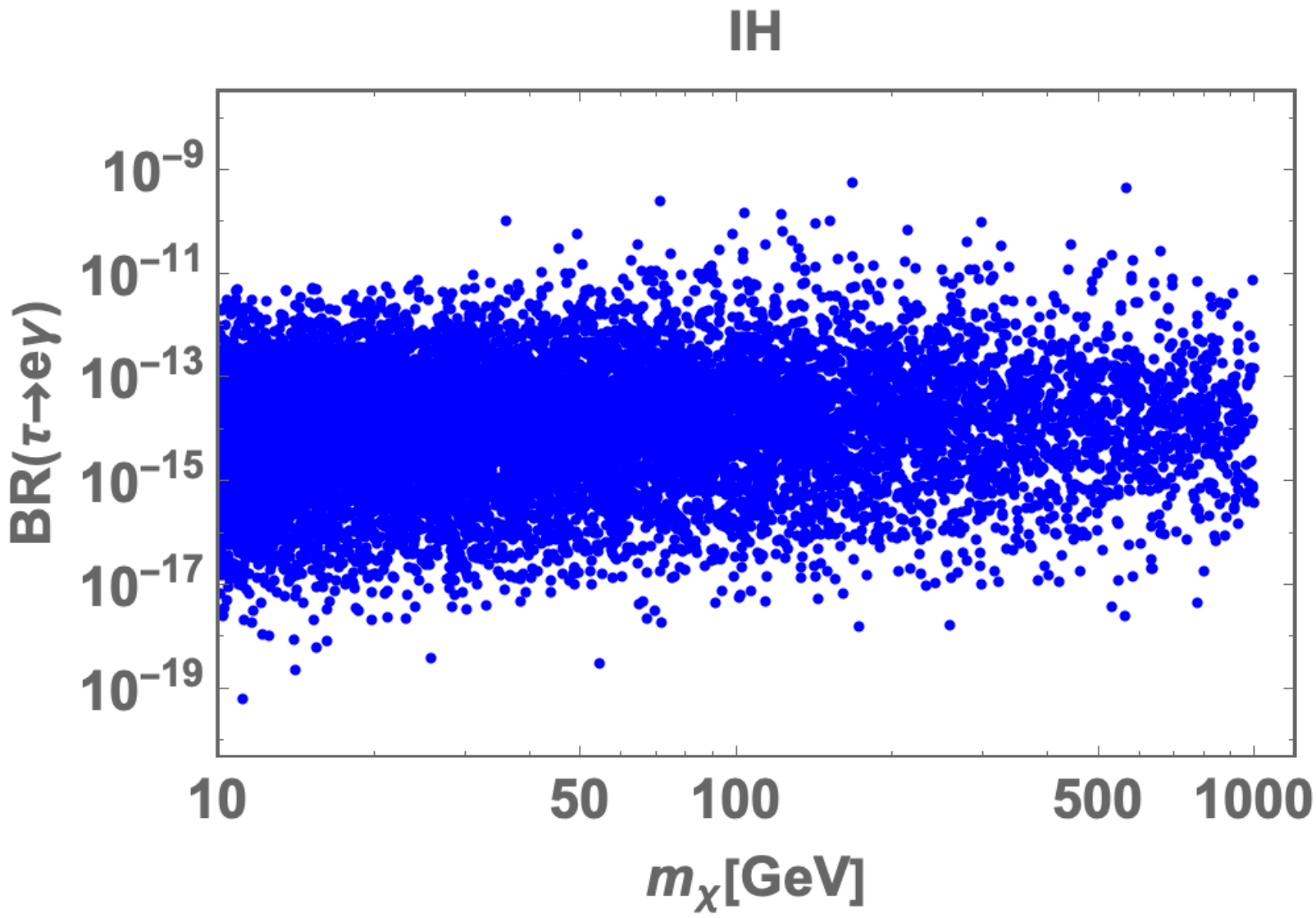}
\includegraphics[width=53mm]{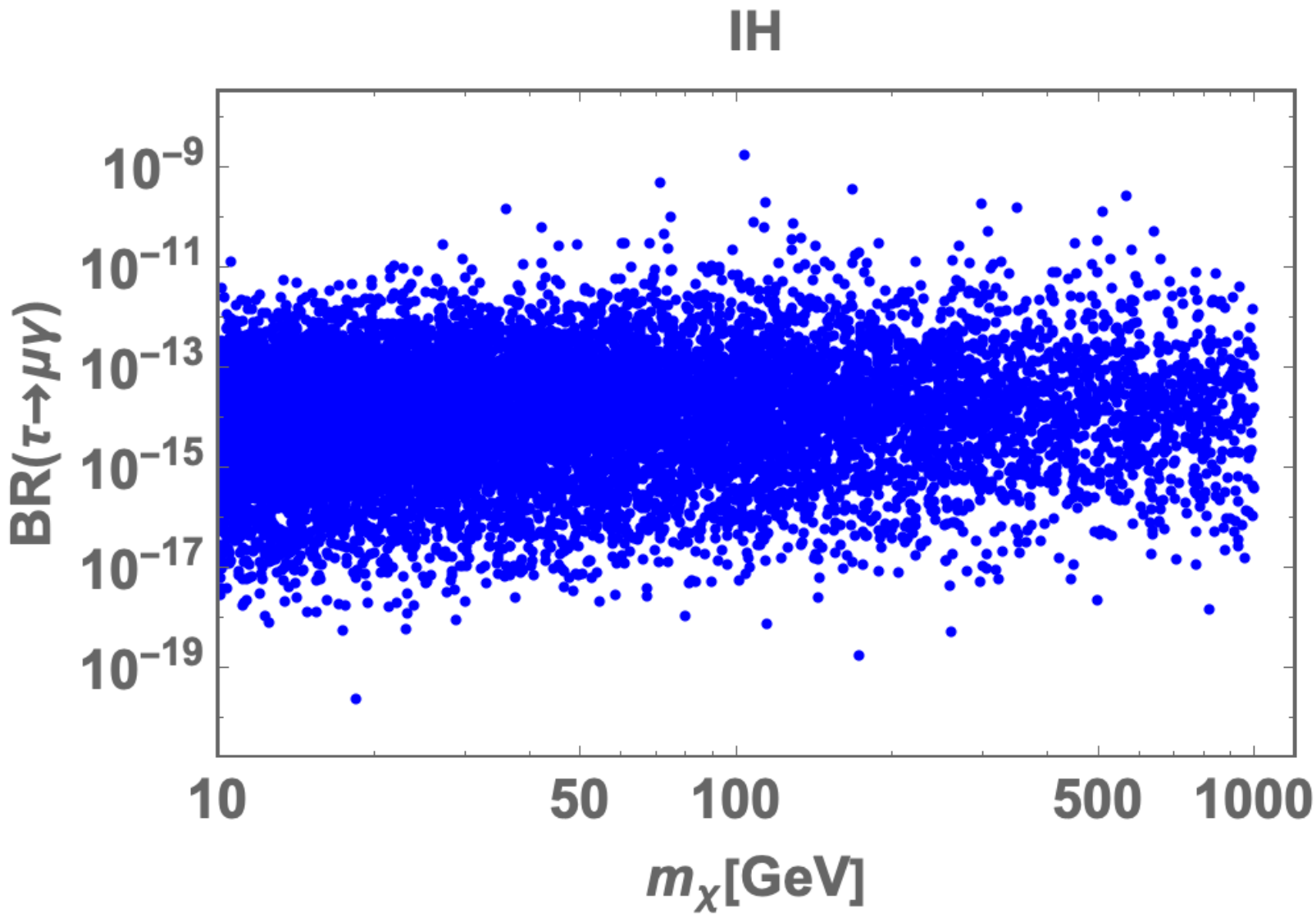}
\caption{Allowed regions for BR($\mu\to e\gamma$)(left), BR($\tau\to e\gamma$)(center), BR($\tau\to \mu\gamma$)(right) in terms of DM mass $N_{R_1}$.} 
\label{fig:ih2}
\end{center}\end{figure}
%
Fig.~\ref{fig:ih2} shows three LFVs; BR($\mu\to e\gamma$)(left), BR($\tau\to e\gamma$)(center), BR($\tau\to \mu\gamma$)(right) in terms of DM mass $N_{R_1}$.
These figures imply that the stringent constraint also comes from  BR($\mu\to e\gamma$) that is well-tested near future.

\section{Summary and discussion}
\label{sec:IV}
We have proposed a radiative lepton seesaw model in the ${\mathbb Z}_3$ Tambara-Yamagami fusion rule and gauged $B-L$ symmetry.
{The symmetry forbids tree-level lepton masses while it is induced at one-loop level via dynamical breaking of the non-invertible selection rules.}
The heavier right-handed neutral fermions ($N_R$) simultaneously contribute to the masses for charged leptons and neutrinos, and we have found that three families of $N_R$ are necessary.
On the other hand, we have introduced a gauged $B-L$ symmetry to preserve the lepton seesaw mechanism. 
As a result, this gauged symmetry have not only assured the three families of $N_R$ from view point of chiral gauge anomaly cancellation, but also provided a dominant annihilation cross section in order to explain the correct relic density of dark matter {through the Higgs portal}. 
Finally, we have shown several numerical results satisfying the neutrino oscillation data, lepton flavor violations, muon $g-2$, and the relic density of dark matter.
\section*{Acknowledgments}
The work was supported by the Fundamental Research Funds for the Central Universities (T.~N.).

\bibliography{ctma4.bib}

\begin{thebibliography}{46}
\expandafter\ifx\csname natexlab\endcsname\relax\def\natexlab#1{#1}\fi
\expandafter\ifx\csname bibnamefont\endcsname\relax
  \def\bibnamefont#1{#1}\fi
\expandafter\ifx\csname bibfnamefont\endcsname\relax
  \def\bibfnamefont#1{#1}\fi
\expandafter\ifx\csname citenamefont\endcsname\relax
  \def\citenamefont#1{#1}\fi
\expandafter\ifx\csname url\endcsname\relax
  \def\url#1{\texttt{#1}}\fi
\expandafter\ifx\csname urlprefix\endcsname\relax\def\urlprefix{URL }\fi
\providecommand{\bibinfo}[2]{#2}
\providecommand{\eprint}[2][]{\url{#2}}

\bibitem[{\citenamefont{Choi et~al.}(2022)\citenamefont{Choi, Lam, and
  Shao}}]{Choi:2022jqy}
\bibinfo{author}{\bibfnamefont{Y.}~\bibnamefont{Choi}},
  \bibinfo{author}{\bibfnamefont{H.~T.} \bibnamefont{Lam}}, \bibnamefont{and}
  \bibinfo{author}{\bibfnamefont{S.-H.} \bibnamefont{Shao}},
  \bibinfo{journal}{Phys. Rev. Lett.} \textbf{\bibinfo{volume}{129}},
  \bibinfo{pages}{161601} (\bibinfo{year}{2022}), \eprint{2205.05086}.

\bibitem[{\citenamefont{Cordova
  et~al.}(2024{\natexlab{a}})\citenamefont{Cordova, Hong, Koren, and
  Ohmori}}]{Cordova:2022fhg}
\bibinfo{author}{\bibfnamefont{C.}~\bibnamefont{Cordova}},
  \bibinfo{author}{\bibfnamefont{S.}~\bibnamefont{Hong}},
  \bibinfo{author}{\bibfnamefont{S.}~\bibnamefont{Koren}}, \bibnamefont{and}
  \bibinfo{author}{\bibfnamefont{K.}~\bibnamefont{Ohmori}},
  \bibinfo{journal}{Phys. Rev. X} \textbf{\bibinfo{volume}{14}},
  \bibinfo{pages}{031033} (\bibinfo{year}{2024}{\natexlab{a}}),
  \eprint{2211.07639}.

\bibitem[{\citenamefont{Cordova and Ohmori}(2023)}]{Cordova:2022ieu}
\bibinfo{author}{\bibfnamefont{C.}~\bibnamefont{Cordova}} \bibnamefont{and}
  \bibinfo{author}{\bibfnamefont{K.}~\bibnamefont{Ohmori}},
  \bibinfo{journal}{Phys. Rev. X} \textbf{\bibinfo{volume}{13}},
  \bibinfo{pages}{011034} (\bibinfo{year}{2023}), \eprint{2205.06243}.

\bibitem[{\citenamefont{Cordova
  et~al.}(2024{\natexlab{b}})\citenamefont{Cordova, Hong, and
  Koren}}]{Cordova:2024ypu}
\bibinfo{author}{\bibfnamefont{C.}~\bibnamefont{Cordova}},
  \bibinfo{author}{\bibfnamefont{S.}~\bibnamefont{Hong}}, \bibnamefont{and}
  \bibinfo{author}{\bibfnamefont{S.}~\bibnamefont{Koren}}
  (\bibinfo{year}{2024}{\natexlab{b}}), \eprint{2402.12453}.

\bibitem[{\citenamefont{Kobayashi et~al.}(2024)\citenamefont{Kobayashi, Otsuka,
  and Tanimoto}}]{Kobayashi:2024cvp}
\bibinfo{author}{\bibfnamefont{T.}~\bibnamefont{Kobayashi}},
  \bibinfo{author}{\bibfnamefont{H.}~\bibnamefont{Otsuka}}, \bibnamefont{and}
  \bibinfo{author}{\bibfnamefont{M.}~\bibnamefont{Tanimoto}},
  \bibinfo{journal}{JHEP} \textbf{\bibinfo{volume}{12}}, \bibinfo{pages}{117}
  (\bibinfo{year}{2024}), \eprint{2409.05270}.

\bibitem[{\citenamefont{Kobayashi and Otsuka}(2024)}]{Kobayashi:2024yqq}
\bibinfo{author}{\bibfnamefont{T.}~\bibnamefont{Kobayashi}} \bibnamefont{and}
  \bibinfo{author}{\bibfnamefont{H.}~\bibnamefont{Otsuka}},
  \bibinfo{journal}{JHEP} \textbf{\bibinfo{volume}{11}}, \bibinfo{pages}{120}
  (\bibinfo{year}{2024}), \eprint{2408.13984}.

\bibitem[{\citenamefont{Kobayashi
  et~al.}(2025{\natexlab{a}})\citenamefont{Kobayashi, Nishioka, Otsuka, and
  Tanimoto}}]{Kobayashi:2025znw}
\bibinfo{author}{\bibfnamefont{T.}~\bibnamefont{Kobayashi}},
  \bibinfo{author}{\bibfnamefont{Y.}~\bibnamefont{Nishioka}},
  \bibinfo{author}{\bibfnamefont{H.}~\bibnamefont{Otsuka}}, \bibnamefont{and}
  \bibinfo{author}{\bibfnamefont{M.}~\bibnamefont{Tanimoto}},
  \bibinfo{journal}{JHEP} \textbf{\bibinfo{volume}{05}}, \bibinfo{pages}{177}
  (\bibinfo{year}{2025}{\natexlab{a}}), \eprint{2503.09966}.

\bibitem[{\citenamefont{Suzuki and Xu}(2025)}]{Suzuki:2025oov}
\bibinfo{author}{\bibfnamefont{M.}~\bibnamefont{Suzuki}} \bibnamefont{and}
  \bibinfo{author}{\bibfnamefont{L.-X.} \bibnamefont{Xu}}
  (\bibinfo{year}{2025}), \eprint{2503.19964}.

\bibitem[{\citenamefont{Liang and Yanagida}(2025)}]{Liang:2025dkm}
\bibinfo{author}{\bibfnamefont{Q.}~\bibnamefont{Liang}} \bibnamefont{and}
  \bibinfo{author}{\bibfnamefont{T.~T.} \bibnamefont{Yanagida}}
  (\bibinfo{year}{2025}), \eprint{2505.05142}.

\bibitem[{\citenamefont{Kobayashi
  et~al.}(2025{\natexlab{b}})\citenamefont{Kobayashi, Otsuka, Tanimoto, and
  Uchida}}]{Kobayashi:2025ldi}
\bibinfo{author}{\bibfnamefont{T.}~\bibnamefont{Kobayashi}},
  \bibinfo{author}{\bibfnamefont{H.}~\bibnamefont{Otsuka}},
  \bibinfo{author}{\bibfnamefont{M.}~\bibnamefont{Tanimoto}}, \bibnamefont{and}
  \bibinfo{author}{\bibfnamefont{H.}~\bibnamefont{Uchida}}
  (\bibinfo{year}{2025}{\natexlab{b}}), \eprint{2505.07262}.

\bibitem[{\citenamefont{Tatsuo~Kobayashi}(2025)}]{Kobayashi:2025lar}
\bibinfo{author}{\bibfnamefont{H.~O. R.~S.} \bibnamefont{Tatsuo~Kobayashi},
  \bibfnamefont{Hironobu~Mita}} (\bibinfo{year}{2025}), \eprint{2506.10241}.

\bibitem[{\citenamefont{Branco et~al.}(1989)\citenamefont{Branco, Lavoura, and
  Mota}}]{Branco:1988iq}
\bibinfo{author}{\bibfnamefont{G.~C.} \bibnamefont{Branco}},
  \bibinfo{author}{\bibfnamefont{L.}~\bibnamefont{Lavoura}}, \bibnamefont{and}
  \bibinfo{author}{\bibfnamefont{F.}~\bibnamefont{Mota}},
  \bibinfo{journal}{Phys. Rev. D} \textbf{\bibinfo{volume}{39}},
  \bibinfo{pages}{3443} (\bibinfo{year}{1989}).

\bibitem[{\citenamefont{Altarelli and Feruglio}(2010)}]{Altarelli:2010gt}
\bibinfo{author}{\bibfnamefont{G.}~\bibnamefont{Altarelli}} \bibnamefont{and}
  \bibinfo{author}{\bibfnamefont{F.}~\bibnamefont{Feruglio}},
  \bibinfo{journal}{Rev. Mod. Phys.} \textbf{\bibinfo{volume}{82}},
  \bibinfo{pages}{2701} (\bibinfo{year}{2010}), \eprint{1002.0211}.

\bibitem[{\citenamefont{Ishimori et~al.}(2010)\citenamefont{Ishimori,
  Kobayashi, Ohki, Shimizu, Okada, and Tanimoto}}]{Ishimori:2010au}
\bibinfo{author}{\bibfnamefont{H.}~\bibnamefont{Ishimori}},
  \bibinfo{author}{\bibfnamefont{T.}~\bibnamefont{Kobayashi}},
  \bibinfo{author}{\bibfnamefont{H.}~\bibnamefont{Ohki}},
  \bibinfo{author}{\bibfnamefont{Y.}~\bibnamefont{Shimizu}},
  \bibinfo{author}{\bibfnamefont{H.}~\bibnamefont{Okada}}, \bibnamefont{and}
  \bibinfo{author}{\bibfnamefont{M.}~\bibnamefont{Tanimoto}},
  \bibinfo{journal}{Prog. Theor. Phys. Suppl.} \textbf{\bibinfo{volume}{183}},
  \bibinfo{pages}{1} (\bibinfo{year}{2010}), \eprint{1003.3552}.

\bibitem[{\citenamefont{Kobayashi et~al.}(2022)\citenamefont{Kobayashi, Ohki,
  Okada, Shimizu, and Tanimoto}}]{Kobayashi:2022moq}
\bibinfo{author}{\bibfnamefont{T.}~\bibnamefont{Kobayashi}},
  \bibinfo{author}{\bibfnamefont{H.}~\bibnamefont{Ohki}},
  \bibinfo{author}{\bibfnamefont{H.}~\bibnamefont{Okada}},
  \bibinfo{author}{\bibfnamefont{Y.}~\bibnamefont{Shimizu}}, \bibnamefont{and}
  \bibinfo{author}{\bibfnamefont{M.}~\bibnamefont{Tanimoto}},
  \emph{\bibinfo{title}{{An Introduction to Non-Abelian Discrete Symmetries for
  Particle Physicists}}} (\bibinfo{year}{2022}), ISBN
  \bibinfo{isbn}{978-3-662-64678-6, 978-3-662-64679-3}.

\bibitem[{\citenamefont{Ishimori et~al.}(2012)\citenamefont{Ishimori,
  Kobayashi, Ohki, Okada, Shimizu, and Tanimoto}}]{Ishimori:2012zz}
\bibinfo{author}{\bibfnamefont{H.}~\bibnamefont{Ishimori}},
  \bibinfo{author}{\bibfnamefont{T.}~\bibnamefont{Kobayashi}},
  \bibinfo{author}{\bibfnamefont{H.}~\bibnamefont{Ohki}},
  \bibinfo{author}{\bibfnamefont{H.}~\bibnamefont{Okada}},
  \bibinfo{author}{\bibfnamefont{Y.}~\bibnamefont{Shimizu}}, \bibnamefont{and}
  \bibinfo{author}{\bibfnamefont{M.}~\bibnamefont{Tanimoto}},
  \emph{\bibinfo{title}{{An introduction to non-Abelian discrete symmetries for
  particle physicists}}}, vol. \bibinfo{volume}{858} (\bibinfo{year}{2012}).

\bibitem[{\citenamefont{Heckman et~al.}(2024)\citenamefont{Heckman, McNamara,
  Montero, Sharon, Vafa, and Valenzuela}}]{Heckman:2024obe}
\bibinfo{author}{\bibfnamefont{J.~J.} \bibnamefont{Heckman}},
  \bibinfo{author}{\bibfnamefont{J.}~\bibnamefont{McNamara}},
  \bibinfo{author}{\bibfnamefont{M.}~\bibnamefont{Montero}},
  \bibinfo{author}{\bibfnamefont{A.}~\bibnamefont{Sharon}},
  \bibinfo{author}{\bibfnamefont{C.}~\bibnamefont{Vafa}}, \bibnamefont{and}
  \bibinfo{author}{\bibfnamefont{I.}~\bibnamefont{Valenzuela}}
  (\bibinfo{year}{2024}), \eprint{2402.00118}.

\bibitem[{\citenamefont{Kaidi et~al.}(2024)\citenamefont{Kaidi, Tachikawa, and
  Zhang}}]{Kaidi:2024wio}
\bibinfo{author}{\bibfnamefont{J.}~\bibnamefont{Kaidi}},
  \bibinfo{author}{\bibfnamefont{Y.}~\bibnamefont{Tachikawa}},
  \bibnamefont{and} \bibinfo{author}{\bibfnamefont{H.~Y.} \bibnamefont{Zhang}}
  (\bibinfo{year}{2024}), \eprint{2402.00105}.

\bibitem[{\citenamefont{Kobayashi
  et~al.}(2025{\natexlab{c}})\citenamefont{Kobayashi, Okada, and
  Otsuka}}]{Kobayashi:2025cwx}
\bibinfo{author}{\bibfnamefont{T.}~\bibnamefont{Kobayashi}},
  \bibinfo{author}{\bibfnamefont{H.}~\bibnamefont{Okada}}, \bibnamefont{and}
  \bibinfo{author}{\bibfnamefont{H.}~\bibnamefont{Otsuka}}
  (\bibinfo{year}{2025}{\natexlab{c}}), \eprint{2505.14878}.

\bibitem[{\citenamefont{Bhardwaj and Tachikawa}(2018)}]{Bhardwaj:2017xup}
\bibinfo{author}{\bibfnamefont{L.}~\bibnamefont{Bhardwaj}} \bibnamefont{and}
  \bibinfo{author}{\bibfnamefont{Y.}~\bibnamefont{Tachikawa}},
  \bibinfo{journal}{JHEP} \textbf{\bibinfo{volume}{03}}, \bibinfo{pages}{189}
  (\bibinfo{year}{2018}), \eprint{1704.02330}.

\bibitem[{\citenamefont{Dijkgraaf et~al.}(1988)\citenamefont{Dijkgraaf,
  Verlinde, and Verlinde}}]{Dijkgraaf:1987vp}
\bibinfo{author}{\bibfnamefont{R.}~\bibnamefont{Dijkgraaf}},
  \bibinfo{author}{\bibfnamefont{E.~P.} \bibnamefont{Verlinde}},
  \bibnamefont{and} \bibinfo{author}{\bibfnamefont{H.~L.}
  \bibnamefont{Verlinde}}, \bibinfo{journal}{Commun. Math. Phys.}
  \textbf{\bibinfo{volume}{115}}, \bibinfo{pages}{649} (\bibinfo{year}{1988}).

\bibitem[{\citenamefont{Kobayashi et~al.}(2005)\citenamefont{Kobayashi, Raby,
  and Zhang}}]{Kobayashi:2004ya}
\bibinfo{author}{\bibfnamefont{T.}~\bibnamefont{Kobayashi}},
  \bibinfo{author}{\bibfnamefont{S.}~\bibnamefont{Raby}}, \bibnamefont{and}
  \bibinfo{author}{\bibfnamefont{R.-J.} \bibnamefont{Zhang}},
  \bibinfo{journal}{Nucl. Phys. B} \textbf{\bibinfo{volume}{704}},
  \bibinfo{pages}{3} (\bibinfo{year}{2005}), \eprint{hep-ph/0409098}.

\bibitem[{\citenamefont{Kobayashi et~al.}(2007)\citenamefont{Kobayashi, Nilles,
  Ploger, Raby, and Ratz}}]{Kobayashi:2006wq}
\bibinfo{author}{\bibfnamefont{T.}~\bibnamefont{Kobayashi}},
  \bibinfo{author}{\bibfnamefont{H.~P.} \bibnamefont{Nilles}},
  \bibinfo{author}{\bibfnamefont{F.}~\bibnamefont{Ploger}},
  \bibinfo{author}{\bibfnamefont{S.}~\bibnamefont{Raby}}, \bibnamefont{and}
  \bibinfo{author}{\bibfnamefont{M.}~\bibnamefont{Ratz}},
  \bibinfo{journal}{Nucl. Phys. B} \textbf{\bibinfo{volume}{768}},
  \bibinfo{pages}{135} (\bibinfo{year}{2007}), \eprint{hep-ph/0611020}.

\bibitem[{\citenamefont{Beye et~al.}(2014)\citenamefont{Beye, Kobayashi, and
  Kuwakino}}]{Beye:2014nxa}
\bibinfo{author}{\bibfnamefont{F.}~\bibnamefont{Beye}},
  \bibinfo{author}{\bibfnamefont{T.}~\bibnamefont{Kobayashi}},
  \bibnamefont{and} \bibinfo{author}{\bibfnamefont{S.}~\bibnamefont{Kuwakino}},
  \bibinfo{journal}{Phys. Lett. B} \textbf{\bibinfo{volume}{736}},
  \bibinfo{pages}{433} (\bibinfo{year}{2014}), \eprint{1406.4660}.

\bibitem[{\citenamefont{Thorngren and Wang}(2024)}]{Thorngren:2021yso}
\bibinfo{author}{\bibfnamefont{R.}~\bibnamefont{Thorngren}} \bibnamefont{and}
  \bibinfo{author}{\bibfnamefont{Y.}~\bibnamefont{Wang}},
  \bibinfo{journal}{JHEP} \textbf{\bibinfo{volume}{07}}, \bibinfo{pages}{051}
  (\bibinfo{year}{2024}), \eprint{2106.12577}.

\bibitem[{\citenamefont{Dong et~al.}(2025)\citenamefont{Dong, Kobayashi,
  Nishida, Nishimura, and Otsuka}}]{Dong:2025pah}
\bibinfo{author}{\bibfnamefont{J.}~\bibnamefont{Dong}},
  \bibinfo{author}{\bibfnamefont{T.}~\bibnamefont{Kobayashi}},
  \bibinfo{author}{\bibfnamefont{R.}~\bibnamefont{Nishida}},
  \bibinfo{author}{\bibfnamefont{S.}~\bibnamefont{Nishimura}},
  \bibnamefont{and} \bibinfo{author}{\bibfnamefont{H.}~\bibnamefont{Otsuka}}
  (\bibinfo{year}{2025}), \eprint{2504.09773}.

\bibitem[{\citenamefont{Funakoshi et~al.}(2024)\citenamefont{Funakoshi,
  Kobayashi, and Otsuka}}]{Funakoshi:2024uvy}
\bibinfo{author}{\bibfnamefont{S.}~\bibnamefont{Funakoshi}},
  \bibinfo{author}{\bibfnamefont{T.}~\bibnamefont{Kobayashi}},
  \bibnamefont{and} \bibinfo{author}{\bibfnamefont{H.}~\bibnamefont{Otsuka}}
  (\bibinfo{year}{2024}), \eprint{2412.12524}.

\bibitem[{\citenamefont{Nomura and Okada}(2024)}]{Nomura:2024ctl}
\bibinfo{author}{\bibfnamefont{T.}~\bibnamefont{Nomura}} \bibnamefont{and}
  \bibinfo{author}{\bibfnamefont{H.}~\bibnamefont{Okada}}
  (\bibinfo{year}{2024}), \eprint{2409.10912}.

\bibitem[{\citenamefont{Nomura and Okada}(2025)}]{Nomura:2025bph}
\bibinfo{author}{\bibfnamefont{T.}~\bibnamefont{Nomura}} \bibnamefont{and}
  \bibinfo{author}{\bibfnamefont{H.}~\bibnamefont{Okada}}
  (\bibinfo{year}{2025}), \eprint{2503.19251}.

\bibitem[{\citenamefont{Lee et~al.}(2021)\citenamefont{Lee, Song, and
  Yamashita}}]{Lee:2021gnw}
\bibinfo{author}{\bibfnamefont{H.~M.} \bibnamefont{Lee}},
  \bibinfo{author}{\bibfnamefont{J.}~\bibnamefont{Song}}, \bibnamefont{and}
  \bibinfo{author}{\bibfnamefont{K.}~\bibnamefont{Yamashita}},
  \bibinfo{journal}{J. Korean Phys. Soc.} \textbf{\bibinfo{volume}{79}},
  \bibinfo{pages}{1121} (\bibinfo{year}{2021}), \eprint{2110.09942}.

\bibitem[{\citenamefont{Chang et~al.}(2019)\citenamefont{Chang, Lin, Shao,
  Wang, and Yin}}]{Chang:2018iay}
\bibinfo{author}{\bibfnamefont{C.-M.} \bibnamefont{Chang}},
  \bibinfo{author}{\bibfnamefont{Y.-H.} \bibnamefont{Lin}},
  \bibinfo{author}{\bibfnamefont{S.-H.} \bibnamefont{Shao}},
  \bibinfo{author}{\bibfnamefont{Y.}~\bibnamefont{Wang}}, \bibnamefont{and}
  \bibinfo{author}{\bibfnamefont{X.}~\bibnamefont{Yin}},
  \bibinfo{journal}{JHEP} \textbf{\bibinfo{volume}{01}}, \bibinfo{pages}{026}
  (\bibinfo{year}{2019}), \eprint{1802.04445}.

\bibitem[{\citenamefont{Ma}(2006)}]{Ma:2006km}
\bibinfo{author}{\bibfnamefont{E.}~\bibnamefont{Ma}}, \bibinfo{journal}{Phys.
  Rev. D} \textbf{\bibinfo{volume}{73}}, \bibinfo{pages}{077301}
  (\bibinfo{year}{2006}), \eprint{hep-ph/0601225}.

\bibitem[{\citenamefont{Casas and Ibarra}(2001)}]{Casas:2001sr}
\bibinfo{author}{\bibfnamefont{J.~A.} \bibnamefont{Casas}} \bibnamefont{and}
  \bibinfo{author}{\bibfnamefont{A.}~\bibnamefont{Ibarra}},
  \bibinfo{journal}{Nucl. Phys. B} \textbf{\bibinfo{volume}{618}},
  \bibinfo{pages}{171} (\bibinfo{year}{2001}), \eprint{hep-ph/0103065}.

\bibitem[{\citenamefont{Aghanim et~al.}(2020)}]{Planck:2018vyg}
\bibinfo{author}{\bibfnamefont{N.}~\bibnamefont{Aghanim}} \bibnamefont{et~al.}
  (\bibinfo{collaboration}{Planck}), \bibinfo{journal}{Astron. Astrophys.}
  \textbf{\bibinfo{volume}{641}}, \bibinfo{pages}{A6} (\bibinfo{year}{2020}),
  \bibinfo{note}{[Erratum: Astron.Astrophys. 652, C4 (2021)]},
  \eprint{1807.06209}.

\bibitem[{\citenamefont{Adame et~al.}(2024)}]{DESI:2024mwx}
\bibinfo{author}{\bibfnamefont{A.~G.} \bibnamefont{Adame}} \bibnamefont{et~al.}
  (\bibinfo{collaboration}{DESI}) (\bibinfo{year}{2024}), \eprint{2404.03002}.

\bibitem[{\citenamefont{Abe et~al.}(2024)}]{KamLAND-Zen:2024eml}
\bibinfo{author}{\bibfnamefont{S.}~\bibnamefont{Abe}} \bibnamefont{et~al.}
  (\bibinfo{collaboration}{KamLAND-Zen}) (\bibinfo{year}{2024}),
  \eprint{2406.11438}.

\bibitem[{\citenamefont{Aker et~al.}(2025)}]{KATRIN:2024cdt}
\bibinfo{author}{\bibfnamefont{M.}~\bibnamefont{Aker}} \bibnamefont{et~al.}
  (\bibinfo{collaboration}{KATRIN}), \bibinfo{journal}{Science}
  \textbf{\bibinfo{volume}{388}}, \bibinfo{pages}{adq9592}
  (\bibinfo{year}{2025}), \eprint{2406.13516}.

\bibitem[{\citenamefont{Baldini et~al.}(2016)}]{MEG:2016leq}
\bibinfo{author}{\bibfnamefont{A.~M.} \bibnamefont{Baldini}}
  \bibnamefont{et~al.} (\bibinfo{collaboration}{MEG}), \bibinfo{journal}{Eur.
  Phys. J. C} \textbf{\bibinfo{volume}{76}}, \bibinfo{pages}{434}
  (\bibinfo{year}{2016}), \eprint{1605.05081}.

\bibitem[{\citenamefont{Aubert et~al.}(2010)}]{BaBar:2009hkt}
\bibinfo{author}{\bibfnamefont{B.}~\bibnamefont{Aubert}} \bibnamefont{et~al.}
  (\bibinfo{collaboration}{BaBar}), \bibinfo{journal}{Phys. Rev. Lett.}
  \textbf{\bibinfo{volume}{104}}, \bibinfo{pages}{021802}
  (\bibinfo{year}{2010}), \eprint{0908.2381}.

\bibitem[{\citenamefont{Renga}(2018)}]{Renga:2018fpd}
\bibinfo{author}{\bibfnamefont{F.}~\bibnamefont{Renga}}
  (\bibinfo{collaboration}{MEG}), \bibinfo{journal}{Hyperfine Interact.}
  \textbf{\bibinfo{volume}{239}}, \bibinfo{pages}{58} (\bibinfo{year}{2018}),
  \eprint{1811.05921}.

\bibitem[{\citenamefont{Afanaciev et~al.}(2024)}]{MEGII:2023ltw}
\bibinfo{author}{\bibfnamefont{K.}~\bibnamefont{Afanaciev}}
  \bibnamefont{et~al.} (\bibinfo{collaboration}{MEG II}),
  \bibinfo{journal}{Eur. Phys. J. C} \textbf{\bibinfo{volume}{84}},
  \bibinfo{pages}{216} (\bibinfo{year}{2024}), \bibinfo{note}{[Erratum:
  Eur.Phys.J.C 84, 1042 (2024)]}, \eprint{2310.12614}.

\bibitem[{\citenamefont{Aliberti et~al.}(2025)}]{Aliberti:2025beg}
\bibinfo{author}{\bibfnamefont{R.}~\bibnamefont{Aliberti}} \bibnamefont{et~al.}
  (\bibinfo{year}{2025}), \eprint{2505.21476}.

\bibitem[{\citenamefont{Ade et~al.}(2014)}]{Planck:2013pxb}
\bibinfo{author}{\bibfnamefont{P.~A.~R.} \bibnamefont{Ade}}
  \bibnamefont{et~al.} (\bibinfo{collaboration}{Planck}),
  \bibinfo{journal}{Astron. Astrophys.} \textbf{\bibinfo{volume}{571}},
  \bibinfo{pages}{A16} (\bibinfo{year}{2014}), \eprint{1303.5076}.

\bibitem[{\citenamefont{Kajiyama et~al.}(2013)\citenamefont{Kajiyama, Okada,
  and Yagyu}}]{Kajiyama:2013zla}
\bibinfo{author}{\bibfnamefont{Y.}~\bibnamefont{Kajiyama}},
  \bibinfo{author}{\bibfnamefont{H.}~\bibnamefont{Okada}}, \bibnamefont{and}
  \bibinfo{author}{\bibfnamefont{K.}~\bibnamefont{Yagyu}},
  \bibinfo{journal}{Nucl. Phys. B} \textbf{\bibinfo{volume}{874}},
  \bibinfo{pages}{198} (\bibinfo{year}{2013}), \eprint{1303.3463}.

\bibitem[{\citenamefont{Workman et~al.}(2022)}]{ParticleDataGroup:2022pth}
\bibinfo{author}{\bibfnamefont{R.~L.} \bibnamefont{Workman}}
  \bibnamefont{et~al.} (\bibinfo{collaboration}{Particle Data Group}),
  \bibinfo{journal}{PTEP} \textbf{\bibinfo{volume}{2022}},
  \bibinfo{pages}{083C01} (\bibinfo{year}{2022}).

\bibitem[{\citenamefont{Esteban et~al.}(2024)\citenamefont{Esteban,
  Gonzalez-Garcia, Maltoni, Martinez-Soler, Pinheiro, and
  Schwetz}}]{Esteban:2024eli}
\bibinfo{author}{\bibfnamefont{I.}~\bibnamefont{Esteban}},
  \bibinfo{author}{\bibfnamefont{M.~C.} \bibnamefont{Gonzalez-Garcia}},
  \bibinfo{author}{\bibfnamefont{M.}~\bibnamefont{Maltoni}},
  \bibinfo{author}{\bibfnamefont{I.}~\bibnamefont{Martinez-Soler}},
  \bibinfo{author}{\bibfnamefont{J.~a.~P.} \bibnamefont{Pinheiro}},
  \bibnamefont{and} \bibinfo{author}{\bibfnamefont{T.}~\bibnamefont{Schwetz}},
  \bibinfo{journal}{JHEP} \textbf{\bibinfo{volume}{12}}, \bibinfo{pages}{216}
  (\bibinfo{year}{2024}), \eprint{2410.05380}.

\end{thebibliography}
\end{document}